\documentclass[10pt,twoside]{article}

\usepackage{graphicx}
\usepackage{amsmath,amsthm}
\usepackage{amssymb}
\begin{document}
\begin{center}

\textbf{\Large{TWO-LETTER WORDS AND A FUNDAMENTAL HOMOMORPHISM RULING  GEOMETRIC CONTEXTUALITY}}
\vspace{2pt}

\baselineskip 16pt
\large{Michel Planat}

\end{center}

\vspace{48pt}

\baselineskip 10pt

%\footnotesize{$^1$ Mathematical physicist, (b. Pessac, France, 1951).}

\footnotesize{\textit{Address}: Michel Planat, Institut FEMTO-ST, CNRS, 15 B Avenue des Montboucons, F-25033 Besan\c con, France.
\\E-mail: michel.planat@femto-st.fr.
%\\Next biographic entries are optional.}

%\footnotesize{\textit{Fields of interest}:
%Geometry, mathematical crystallography (also ornamental arts, anthropology - non-professional interests in parentheses).}

%\footnotesize{\textit{Awards}: Symmetry Award, 1987; Dissymmetry Medal, 1989.}

%\footnotesize{\textit{Publications and/or Exhibitions}:
%List your main symmetry-related publications/exhibitions in chronological order,
%following the conventions of the references.
%Please, give no more than the five most characteristic items here.}

\vspace{2pt}

%\vspace{2pt}
%$^*$ \footnotesize{Corresponding author}

\vspace{24pt}

\normalsize

\textbf{Abstract}:
\textit{
It has recently been recognized by the author that the quantum contextuality paradigm may be formulated in terms of the properties of some subgroups of the two-letter free group $G$ and their corresponding point-line incidence geometry $\mathcal{G}$. I introduce a fundamental homomorphism $f$ mapping the (infinitely many) words of G to the permutations ruling the symmetries of $\mathcal{G}$. The substructure of $f$ is revealing the essence of geometric contextuality in a straightforward way.
}

\vspace{12pt}


% Replace these PACS codes by your own appropriate ones
% or suppress the line if these codes are inappropriate
%\textbf{PACS}: 11.30.Qc, 24.80.+y, 33.15.Bh, 61.50.Ah, 87.15.B-, 78.20.Ek

% Replace these AMS Math Subject Classification codes by your own appropriate ones
% or suppress the line if these codes are inappropriate

\vspace{12pt}
\textbf{\large{Introduction}}

You would like to describe the world from only two letters $a$ and $b$, with relations \lq rels' between them, and reveal the symmetries of the resulting Lego. You perfectly know group theory and that $G=\left\langle a,b|\mbox{rels} \right\rangle$ would have infinitely many words, even after factorizing, like in the word $ab^2(aa^{-1})b^7a^{-1}=ab^9a^{-1}$ (take care that commutativity doesn't hold). Sometimes, I mean for well choosen $\mbox{rels}$ and a subgroup $H$ of finite index $n$ in $G$, you could divide $\infty$ as $\frac{\infty}{n}=\infty$ and so doing you would arrive at structures compatible with $\mbox{rels}$, and picture them as a kind of \lq dessin d'enfant' (this has been recognized by A. Grothendieck (1928-2014) \cite{Groth84} and developed in \cite{PlanatGiorgetti2015}-\cite{PlanatZoology2016}), e.g. for $G=\left\langle a,b|b^2=e \right\rangle$, $e$ the identity of $G$, $n=6$ and a few extra reasonable constraints to be described below, you would draw dessins d'enfants like (i)-(k) in Fig. 1.  

\begin{figure}[ht]
\centering 
\includegraphics[width=4cm]{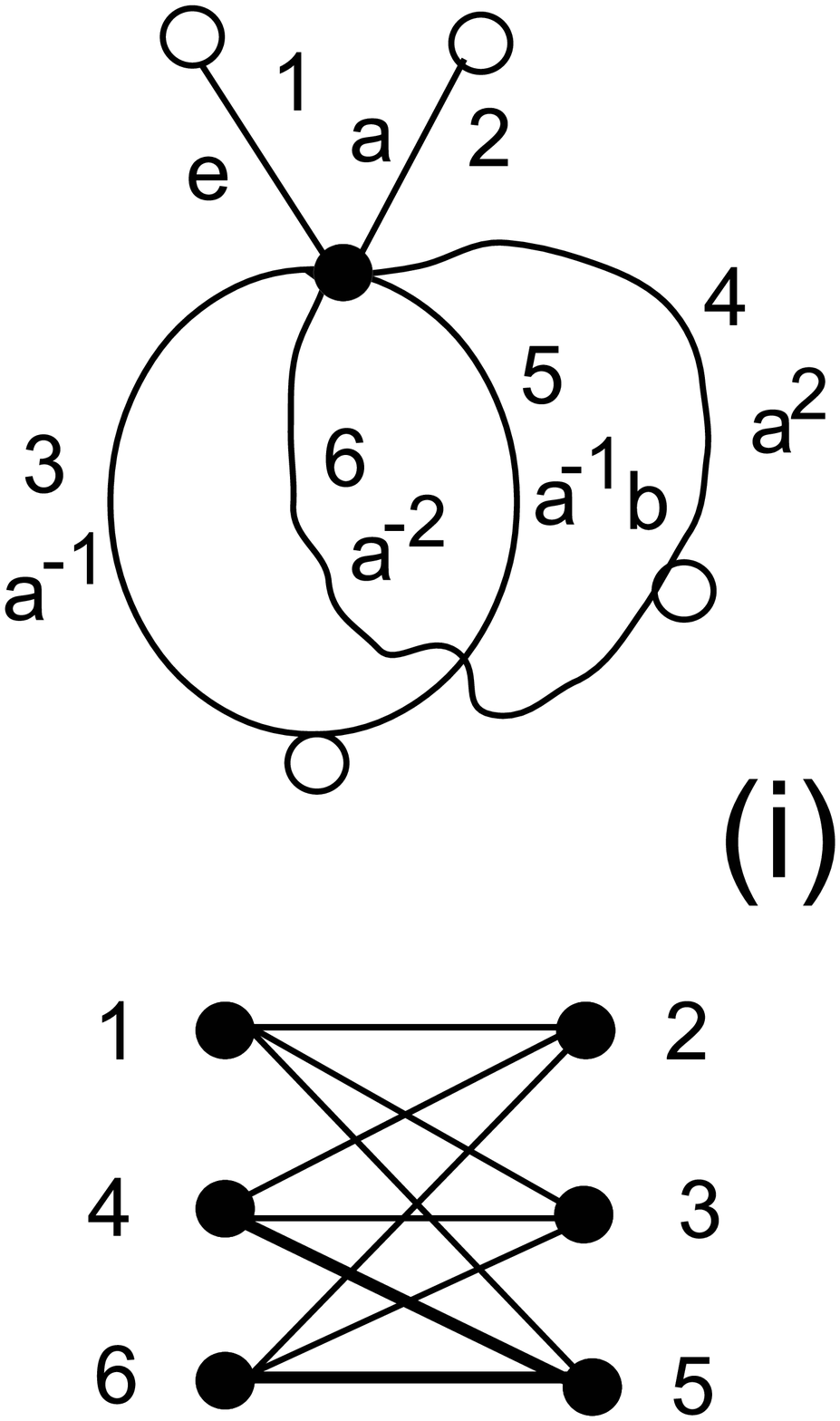}
\includegraphics[width=4cm]{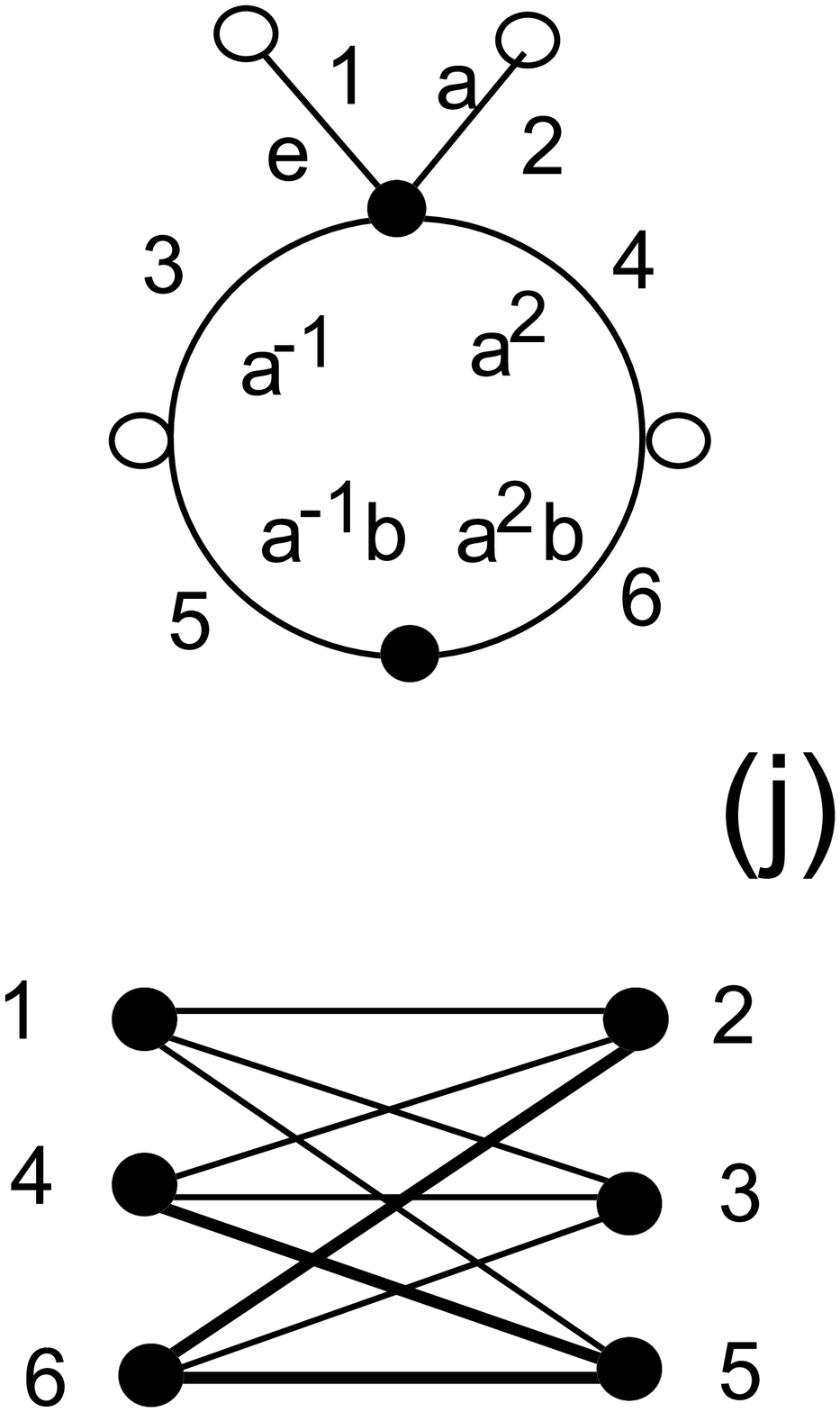}
\includegraphics[width=4cm]{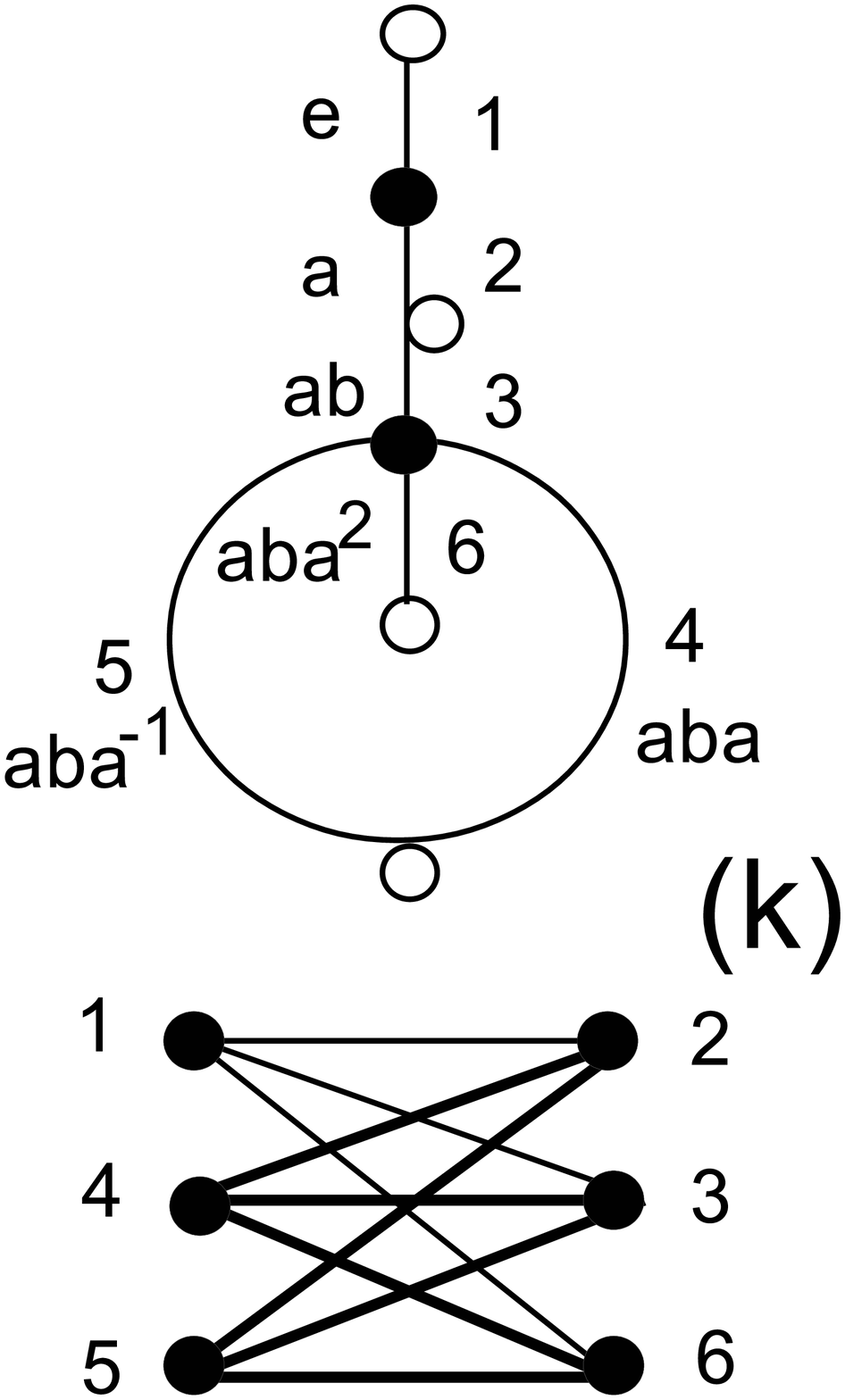}
\caption{Three dessins (i), (j) and (k) of index $6$ stabilizing the bipartite graph $K(3,3)$. Thick edges have non commuting cosets/vertices. Since Axiom 2 of Sec. 1. is not satisfied, dessins (i)-(k) are considered to be non contextual.
 }
\end{figure} 

Such dessins have black and white points connected by edges labelled with elements in the set $S=\{1,..,6\}$ or with short words that are representatives of the (coset) classes you defined while cutting the group $G$ \cite{Planat2015}.
Instead of drawing, you could represent dessins as permutation groups with two generators acting on $S$ as
 $P_{(i)}=\left\langle n|\alpha,\beta\right\rangle=\left\langle 6|(1,2,4,5,6,3),(3,5)(4,6)\right\rangle$ (left of Fig.1), $P_{(j)}=\left\langle 6|(1,2,4,3),(3,5)(4,6)\right\rangle$ (middle of Fig.1) and $P_{(k)}=\left\langle 6|(1,2)(3,4,6,5),(2,3)(4,5)\right\rangle$ (right of Fig.1), all of three isomorphic to the group $P_{36}\cong\mathbb{Z}_3^2 \rtimes \mathbb{Z}_2^2$. In addition, the substructure of the dessins can be summarized as the bipartite graph $K(3,3)$ shown at the bottom of each dessin, whose edges are thin if they have endpoints with commuting cosets and thick otherwise. The latter dichotomy is what we intend to describe below as a fingerprint of \lq geometric contextuality" (see Definition 1 in Sec. 1.).

Axioms for geometric contextuality are in Sec. 1 and results are in Secs 2 to 4. Our conclusion emphasizes the spirit of our approach for the layman.

\section{Geometric contextuality}

All the mathematical 	operations to be described below can be performed with a simple code in Magma \cite{Magma}.

Let us now describe the move from $G$ to its subgroup $H$ as an homomorphism $G \xrightarrow[f] {}P=P_H$ that maps the infinitely many elements of $G$ to the permutations in the finite group $P$ modulo the group of automorphisms of $P$ \cite{Magma}. The homomorphism $f$ may or may not be the same than the natural homomorphism $f_0$ mapping $a \xrightarrow[f_0] {}\alpha$ and $b \xrightarrow[f_0] {}\beta$, that is, sending the generators $a$ and $b$ of the free group to the generators $\alpha$ and $\beta$  of the permutation group $P$ of the dessin.

Further we define the move from the permutation group $P=P_H$ to a graph/geometry $\mathcal{G}$ that fully characterizes $P$. There exists $s\ge 2$ two-point stabilizer subgroups of $P$. Given one of them $P_{\mbox{sub}}$, the edges/lines of $\mathcal{G}$ are defined by the stabilizer subgroups isomorphic (but unequivalent) to $P_{\mbox{sub}}$ \cite{PlanatGiorgetti2015}-\cite{Planat2015}.

{\bf Axiom 1}. About the stabilizer subgroups of $P$ \footnote{The stabilizer subgroup of $G$ at point $x$ is the set $G.x=\{g.x|g \in G\}$.}: the rank $r$ of $P$, alias the number of orbits of the one-point stabilizer subgroup of $P$ {\bf and} the number $s$ of two-point stabilizer subgroups of $P$ (up to isomorphism) should be both at least $3$. 

{\bf Axiom 2}. About the normal closure $N$ of the subgroup $H$ in $G$ (that is, the smallest normal subgroup of $G$ containing $H$): one requires that $N=G$. 

{\bf Definition 1}. First obey Axioms 1 and 2, then a candidate \lq contextual' structure satisfies $f=f_0$ (that is, homomorphisms $f$ and $f_0$ are the same) while $f \ne f_0$ means non-contextuality. In addition, the elected \lq contextual geometry' $\mathcal{G}$ should have the same two-point stabilizer subgroup on a line.

Let us justify these axioms. In \cite{Planat2015}, a graph (or a geometry) $\mathcal{G}$ with edges (or lines) containing vertices (or points) with mutually commuting cosets is called non-contextual, and is called contextual if one edge (or line) has non-commuting cosets. All three graphs in Fig. 1 [the bipartite graph $K(3,3)$] have some edges with non-commuting cosets. Axiom 1 is satisfied (with $r=s=3)$ but Axiom 2 is not, thus such structures are considered to be non-contextual. By the way, this statement fits the physical intuition that, for a two-qubit coordinatization of the bipartite graph $K(3,3)$, Alice spins ($XI,YI,ZI)$ commute with Bob spins ($IX,IY,IZ$); however they cannot be used as a proof (a la Kochen-Specker) of contextuality.

\section{Results for index $n \le 12$. }

\begin{table}[t]
\begin{center}
% \vspace*{-0.4cm}
\begin{tabular}{|l|c|r|c|r|}
\hline 
index & all & non-trivial & contextual & types \\
\hline
6 & 56 & 7 & 0 & -\\
8 & 482 & 37 & 0 & -\\
9 & 1551 & 53 & 12 (4) & Mermin square, $K(3,3)$, Pappus\\
10 & 5916 & 351 & 2(1)& Mermin pentagram\\
12 & 90033 & 3982 & many &$K(4,4,4)$, $[12_6,24_3]_4$ \\
\hline
\end{tabular}
\caption{In column 2, all is the number of conjugacy classes of subgroups of the corresponding index. In column 3, the non-trivial structures with $r$ and $s$ >2 are given, as defined in Axiom 1. In column 3, one finds the number of candidate contextual structures (as defined in Definition 1). For the number inside the parentheses one restricts to (one elects) such geometries $\mathcal{G}$ whose maximum cliques of the collinearity graph have their vertices corresponding to the same stabilizer subgroup of $P$. Column 4 lists the possible types of contextual geometries.}
\end{center}
\end{table}

Our present definition of geometric contextuality is based on a group structure characterized by a quadruple $(G,H,P,\mathcal{G})$. One requires that the normal closure $N$ of the subgroup $H$ of $G$ obeys $N=G$ (Axiom 2), that the symmetry $P$ attached to cosets of $H$ in $G$ is non-trivial (Axiom 1), that the homomorphism $G \xrightarrow[f] {}P$ is unique and that the mapping $P \rightarrow \mathcal{G}$ corresponds to distinguishing the lines of $\mathcal{G}$ by their distinct (but isomorphic) two-point stabilizer subgroups (Definition 1). 
The new criterion allows a fast computation of contextual non-trivial structures up to index $12$, as shown in Table 1.

In \cite{Planat2015}, we defined geometric contextuality in a more fuzzy way as the lack of commutativity of cosets on some lines of $\mathcal{G}$. We get a similar conclusion (found in Table 1) that geometric contextuality starts at index $9$ with Mermin square and that contextual structures are filtered in a very selective way.

Standard contextual structures arising for index $9$ and $10$ are Mermin's geometries \cite{PlanatGiorgetti2015}-\cite{PlanatMM2016}: the $3 \times 3$ square in Fig. 2i and Mermin's pentagram in Fig. 2j (the drawings are taken from Fig. 3 of \cite{Planat2015}).

\begin{figure}
\centering 
\includegraphics[width=4cm]{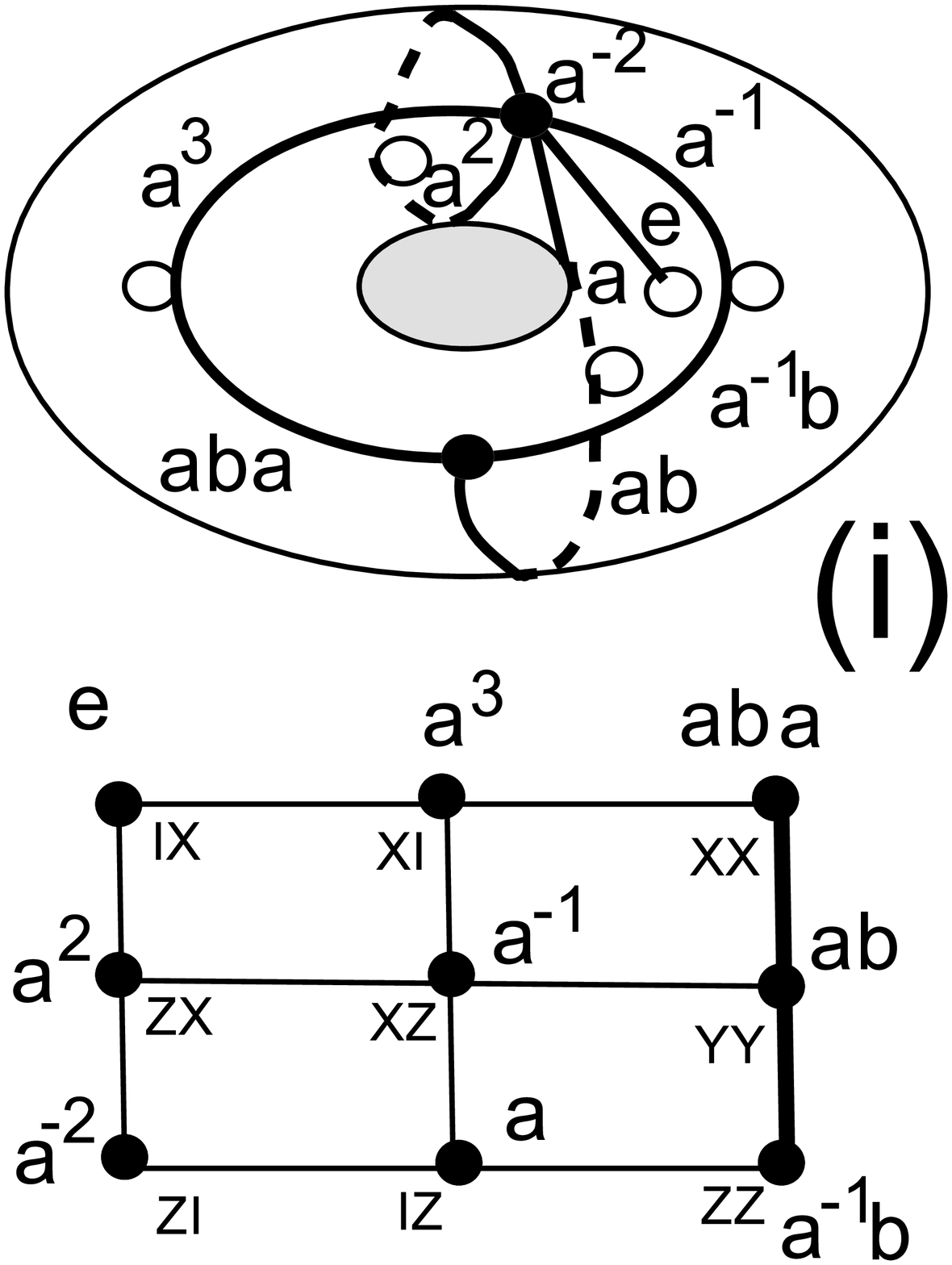}
\includegraphics[width=4cm]{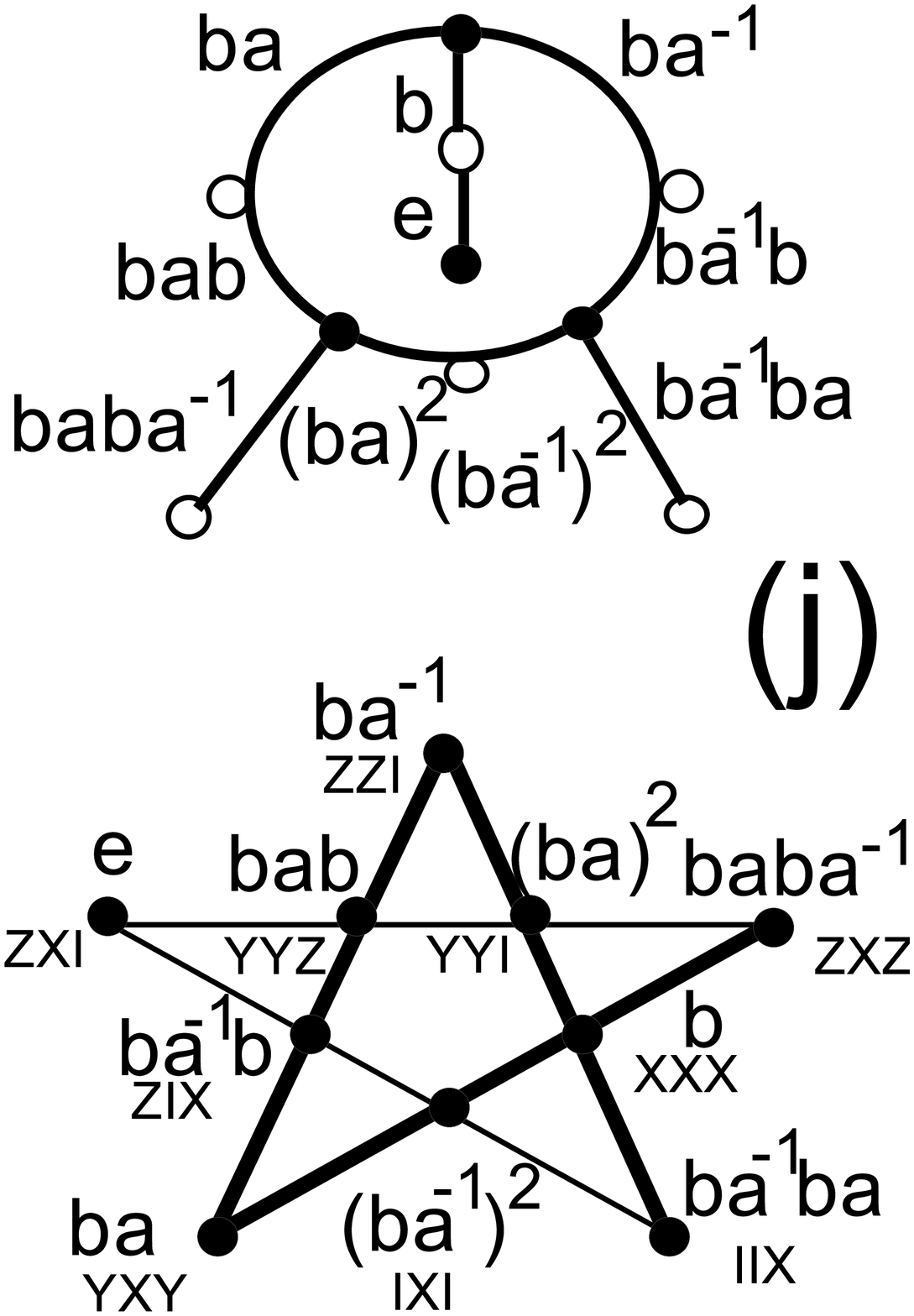}
\caption{(i) The contextual map/dessin (top) stabilizing the Mermin square (bottom) with the corresponding coset labelling. The right hand side column is defective as in the original proof of Kochen-Specker theorem derived for two-qubit coordinates. (j) The contextual map/dessin (top) stabilizing Mermin pentagram (bottom). The thick lines are defective: not all of their cosets are commuting. The lines of the pentagram are given three-qubit coordinates in such a way that the product of operators on a thick line is minus the identity matrix.}
\label{fig2}
\end{figure}

\begin{figure}
\centering 
\includegraphics[width=4cm]{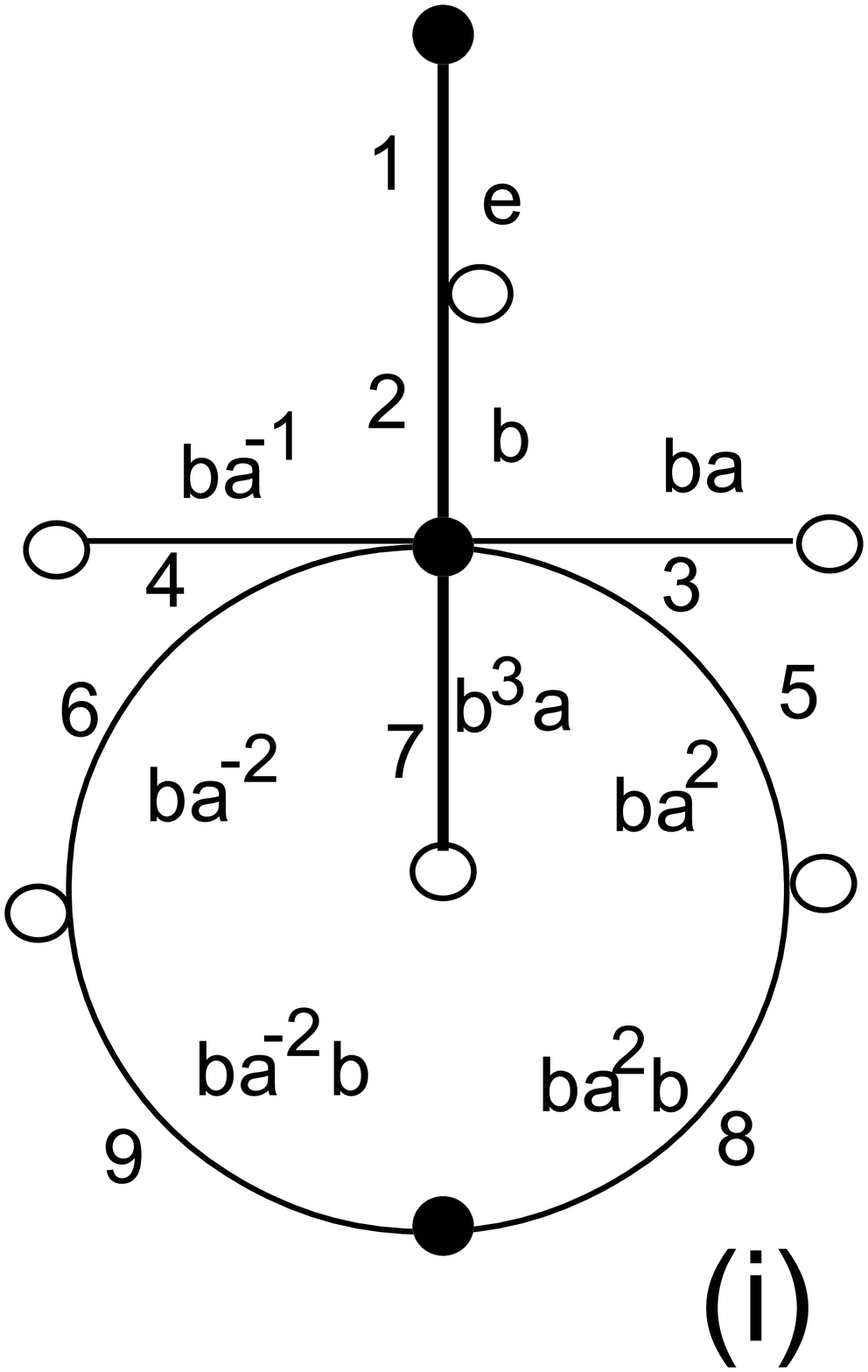}
\includegraphics[width=4cm]{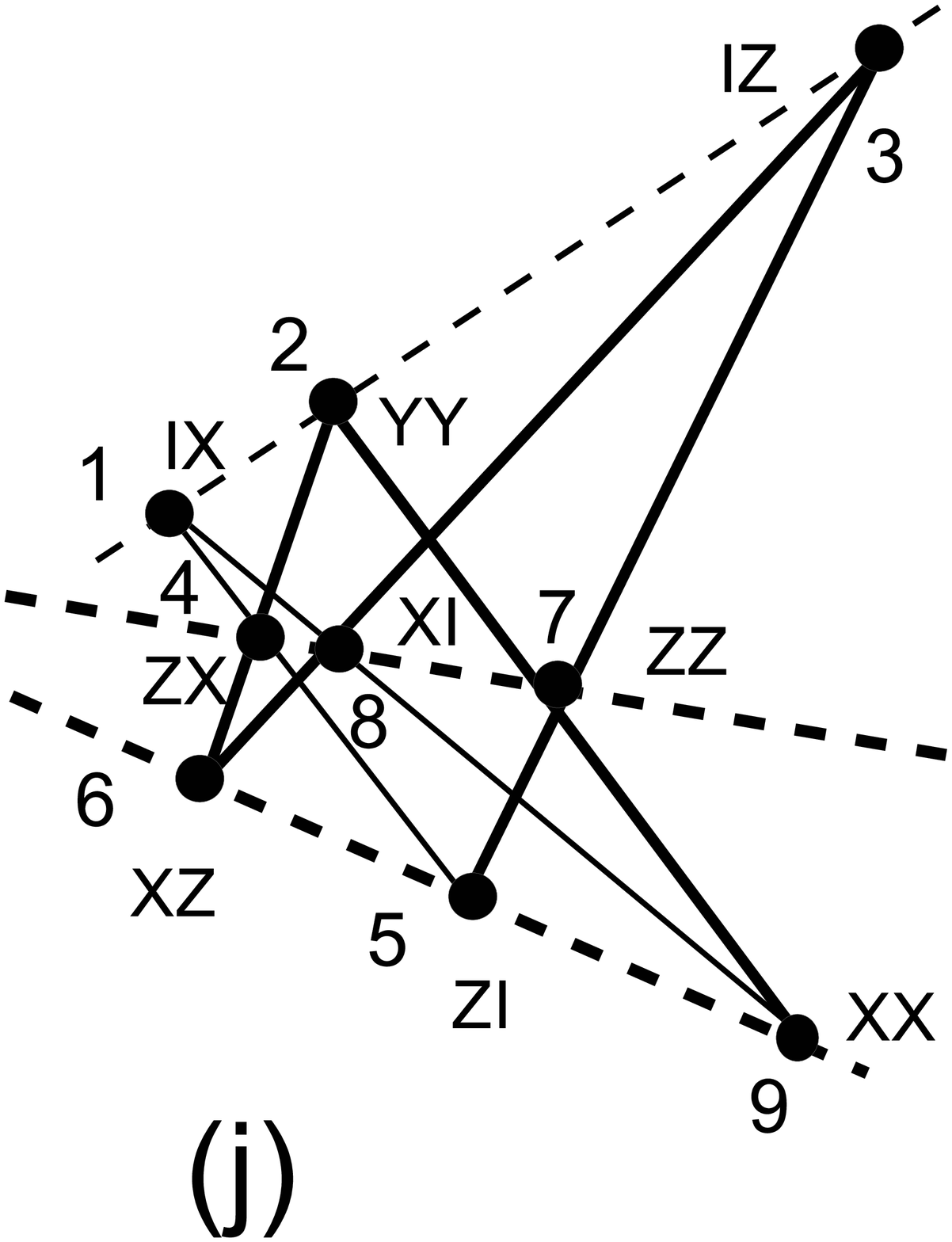}
\caption{The contextual map (i) stabilizing Pappus configuration in (j). The configuration embeds the Mermin configuration (plain lines): dotted lines do not belong to the Mermin square. As before, thick lines correspond to non-commuting cosets.
}
\label{fig3}
\end{figure}

Mermin's square in Fig. 2i is the only non-trivial (from Axiom 2) structure in $G$ with group $P \cong P_{36}$ encountered in Fig. 1.
% and has (contextual) supset structure $G \supset_{36} K$, $G \supset_{9} N=H \supset_{4} K$.
Mermin's pentagram in Fig. 2j is the only non-trivial contextual structure arising with index $10$ and the modular group $\Gamma=\left\langle a,b|a^2=b^3=e \right\rangle$ with group $P \cong A_5$ (the $5$-letter alternating group).

From now on, for the sake of conciseness, we do not explicit any more the coset representatives in the set $S=\{1,..,n\}$ (n the index).

{\bf For index $9$}, there exists a unique dessin (not shown), with permutation group of order $54$, stabilizing the bipartite graph $K(3,3,3)$ in a contextual way (with the $27$ triangles, as lines of $K(3,3,3)$, distinguished by their stabilizer subgroup). There also exist two dessins, with permutation group of order $108$, stabilizing the Pappus configuration. One of the two is pictured in Fig. 3. The plain lines in Fig. 3j illustrate the fact that the Pappus can embed the Mermin square accounted for in Fig. 2i.

The remaining $8$ dessins of index $9$ (see column 4 of Table 1), of permutation group of order bigger than $108$, define the tripartite graph $K(3,3,3)$ from the maximum cliques of the collinearity graph but some of the $27$ triangles fail to correspond to the same stabilizer subgroup of $P$.  

\begin{figure}
\centering 
\includegraphics[width=4cm]{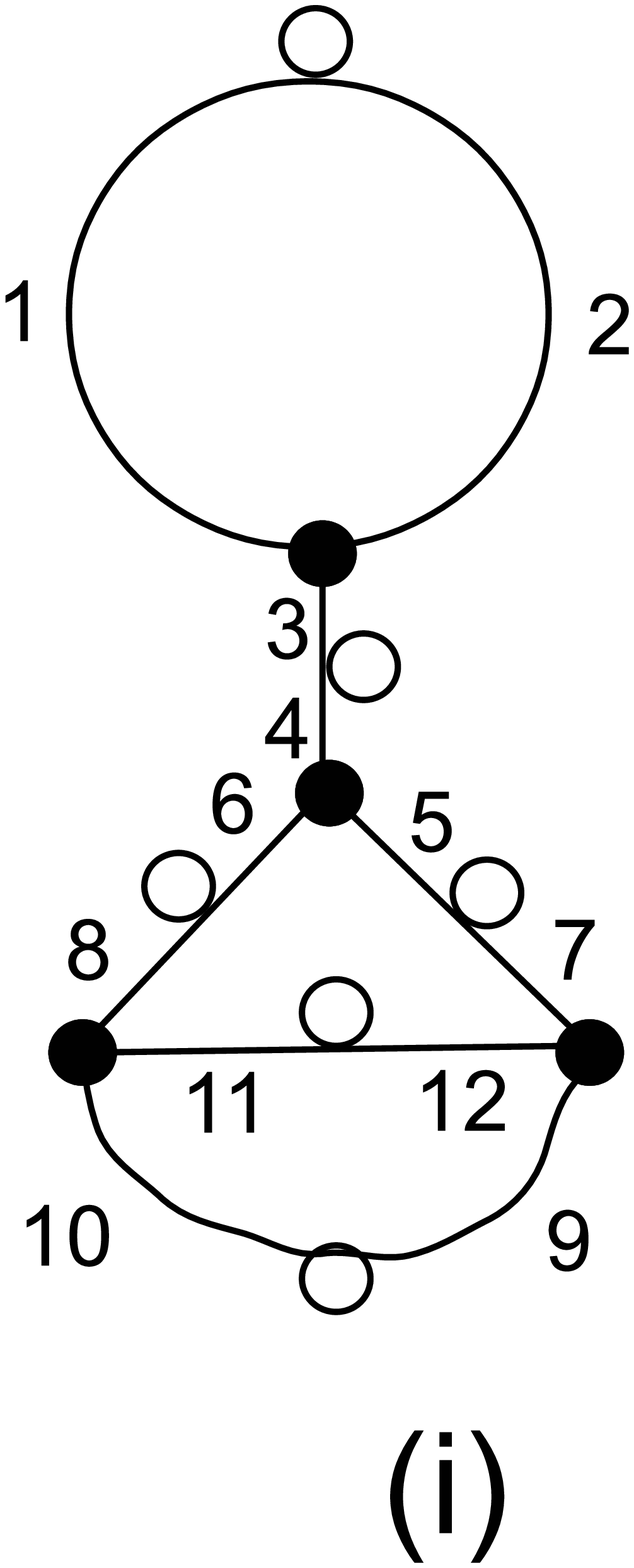}
\includegraphics[width=4cm]{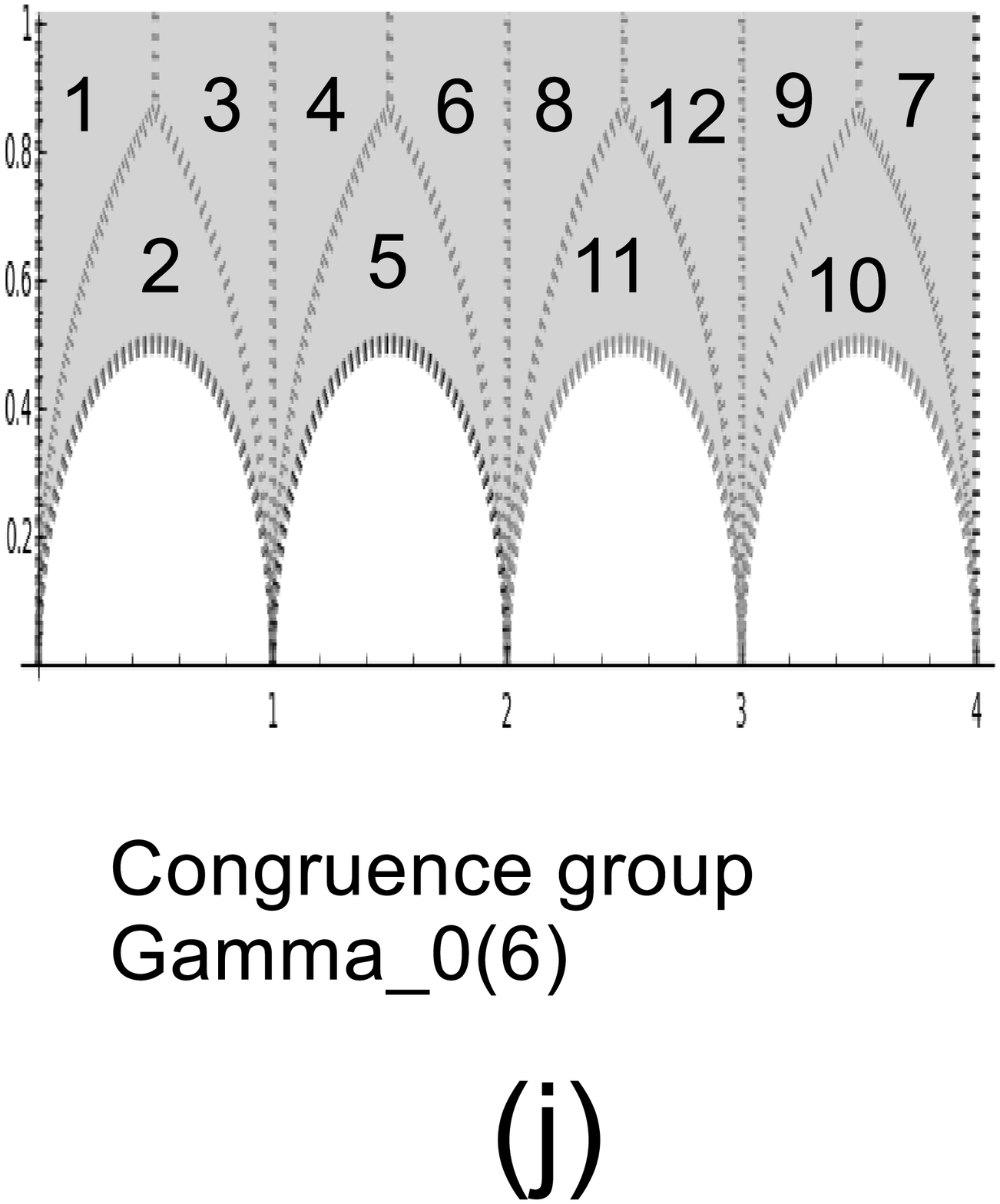}
\includegraphics[width=4cm]{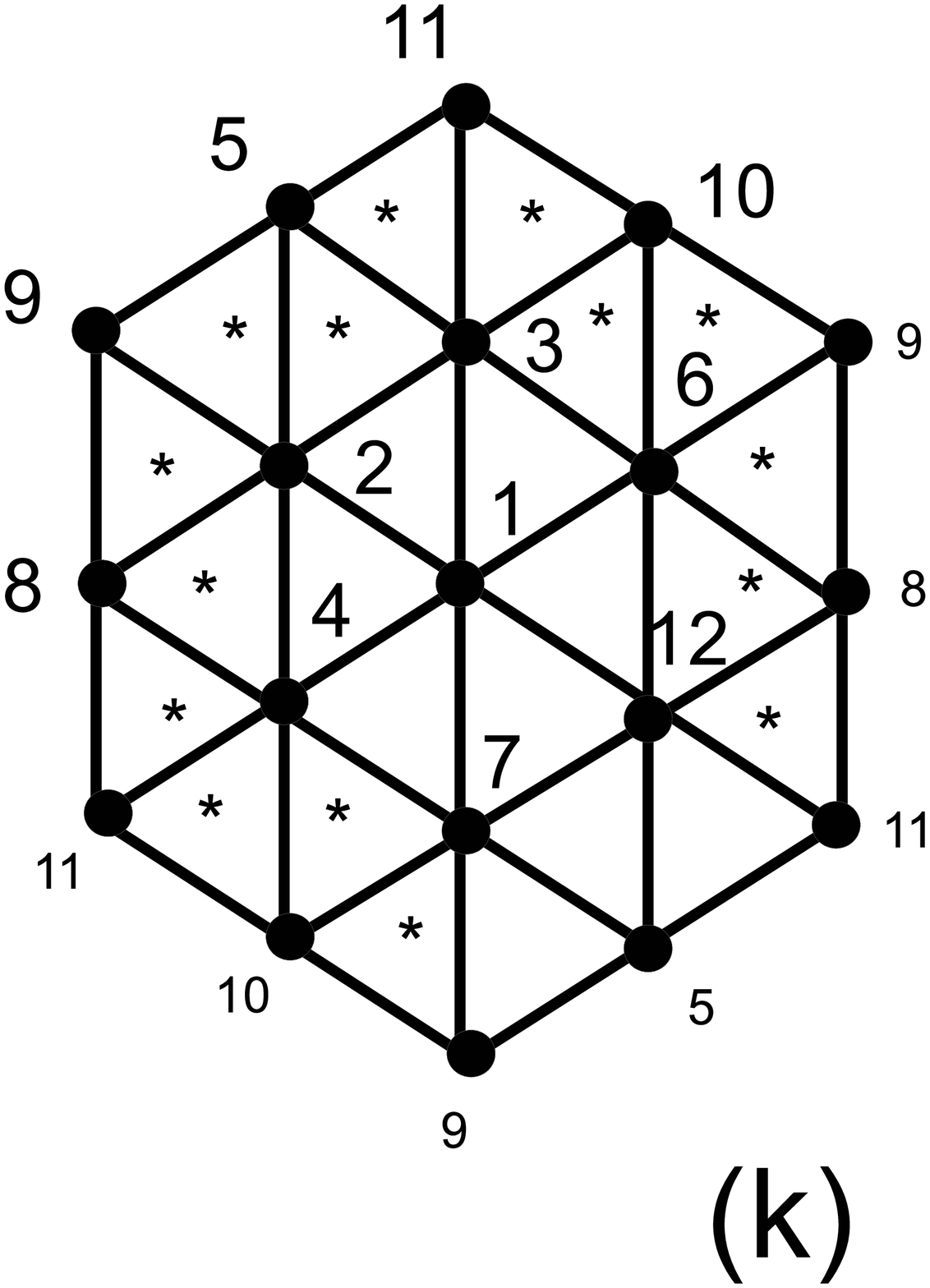}
\caption{The contextual dessin d'enfant (i) with permutation group of order $72$. It corresponds to the congruence subgroup $\Gamma_0(6)$ of modular group $\Gamma$ as shown in (j). The stabilized configuration is in (k). Triangles with a star inside have non-commuting cosets.}
\label{fig4}
\end{figure}

{\bf For index $12$}, there are many cases satisfying the contextual statement of Definition 1. Most of the corresponding dessins stabilize the tripartite graph $K(4,4,4)$ including when the permutation group $P$ is the smallest one, isomorphic to the group $\mathbb{Z}_2 \times S_4$, where $S_4$ is the symmetric group on $4$ letters.

A more exotic configuration is stabilized by the dessin shown in Fig. 4i whose permutation group is of order $72$. As black points are trivalent, the dessin corresponds to a subgroup of the modular group $\Gamma$ that is recognized to be the congruence subgroup $\Gamma_0(6)$ of $\Gamma$ depicted in Fig. 4j. The normalizer of $\Gamma_0(6)$ in $\Gamma$ is the moonshine group $\Gamma_0^+(6)$ \cite{PlanatMoonshine2015}. The configuration $\mathcal{G}=[12_6,24_3]_{(4)}$ is of rank $4$, comprises $12$ points and $24$ lines/triangles with $6$ lines through each point. It is pictured in Fig. 4k. \footnote{The configuration $\mathcal{G}$ is improperly defined as the bipartite graph $K(6,6)$ in \cite{Sebbar2002}-\cite{PlanatMoonshine2015} (table 1) because it has the same number of edges. }. 
The group of automorphisms of $\mathcal{G}$ is isomorphic to $\mathbb{Z}_2^4 \rtimes P_{36}$, where $P_{36}$ was encountered in Fig. 1 as the symmetry group of the Mermin square. The complement of the collinearity graph of $\mathcal{G}$ is the $(3 \times 4)$-grid that physically corresponds to the geometry of the $12$ maximum sets of commuting operators in a qubit-qutrit system \cite{PlanatQubitQutrit2008}. Two points on a line of the grid correspond to maximum sets having one point in common while the triangles in (k) correspond to maximum sets of (three) mutually ubiased bases.

\section{Results for index $n > 12$.}

{\bf For index $15$}, multipartite graphs $K(5,5,5)$ and $K(3,3,3,3,3)$ may be stabilized. Other found configurations are the triangular graph $T(6)$ and $[15_{12},60_3]_{(4)}$ with $60$ lines and rank $r=4$. There also exists a unique pair $(G,H)$ with $G=\left\langle a,b|b^2=a^6=e \right\rangle$ and $P \cong S_6$ (where $S_6$ is the six-letter symmetric group). The dessin $P$ stabilizes the generalized quadrangle of order two $GQ(2,2)$, a model of two-qubit commutation, as pictured in Fig. 5.

\begin{figure}
\centering 
\includegraphics[width=5cm]{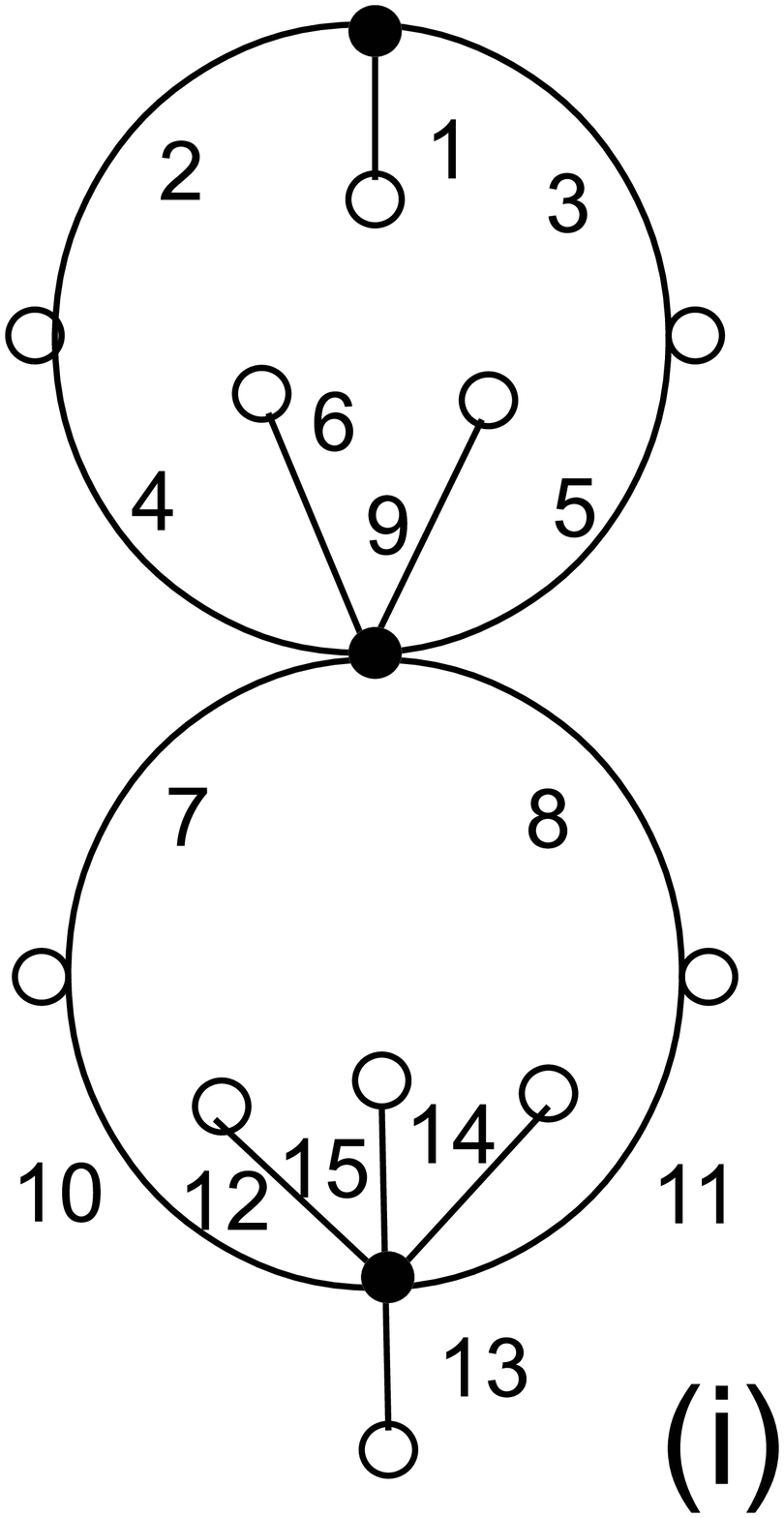}
\includegraphics[width=5cm]{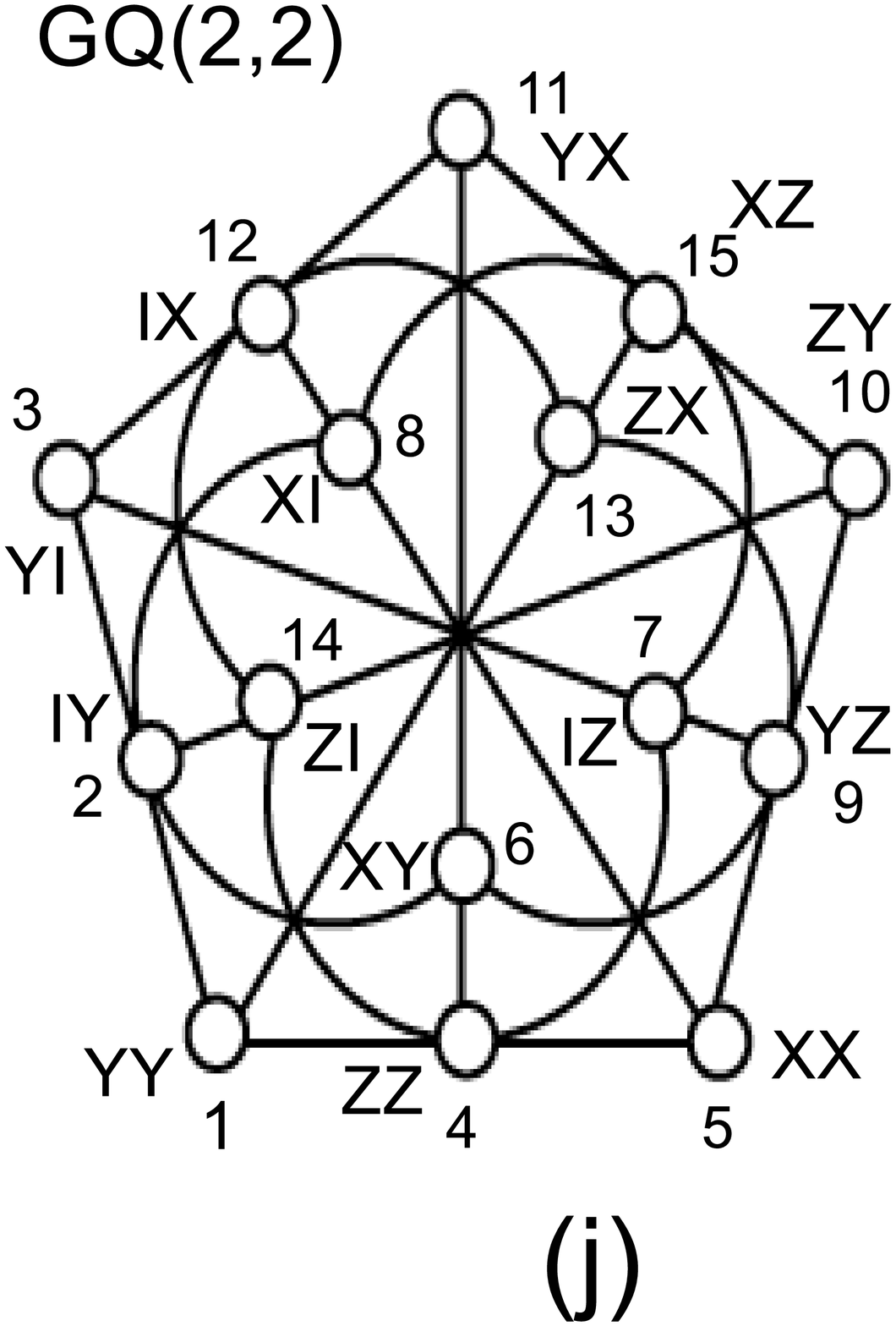}
\caption{The contextual dessin d'enfant (i) with permutation group isomorphic to $S_6$. It stabilizes the generalized quadrangle of order two $[15]_3=GQ(2,2)$ shown in (j). this configuration is a model of a two-qubit system. The list of cosets labelled from $1$ to $15$ is as  follows 
$[e, a, a^{-1}, ab, a^{-1}b, aba, aba^{-1}, b^a, a^{-1}ba^{-1}, aba^{-1}b, a^{-1}bab, aba^{-1}ba, aba^{-1}ba^{-1}, a^{-1}baba^{-1},$           $ aba^{-1}ba^2]$.
 Only lines through the identity element have commuting cosets.}
\label{fig5}
\end{figure}

From now on, we restrict to contextual structures built from the finite representation
$G=\left\langle a,b|b^2=a^3=e \right\rangle$ of the modular group $\Gamma$. 

{\bf For index $16$}, and $G \cong \Gamma$, there exist two subgroups $H$ of $G$ satisfying Definition 1. The first one leads to the multipartite graph $K(4,4,4,4)$ and the second one is associated to the rank $5$ configuration $[16_6,32_3]_{(5)}$ of {\bf Schrikhande graph}, as pictured in Fig. 6.
One can use the software Sage to pass from the finite representation of $H$ to the  in modular representation in $\Gamma$, and the list of small index congruence subgroups of $\Gamma$ in \cite{Cummins} to recognize Fig. 6j as the congruence subgroup of type $8F_0$, of level $8$, with number of elliptic points $(\nu_2,\nu_3)=(4,1)$ and cusps of structure $8^2$.
The Shrikhande graph is shown on Fig. 6k. Each point is at the center of an hexagon and the $32$ lines/triangles are clearly shown in this toric representation.  

\begin{figure}
\centering 
\includegraphics[width=4cm]{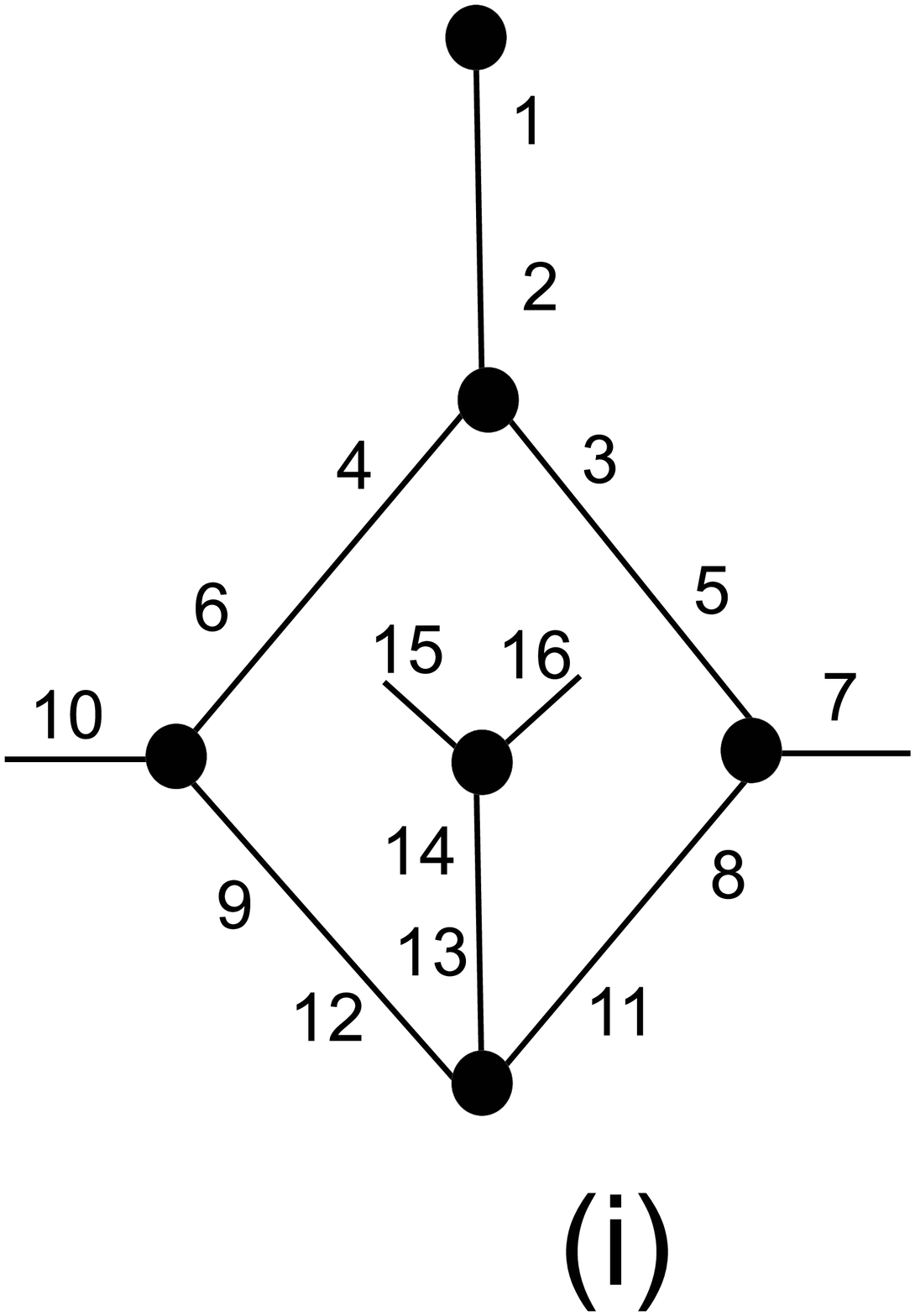}
\includegraphics[width=4cm]{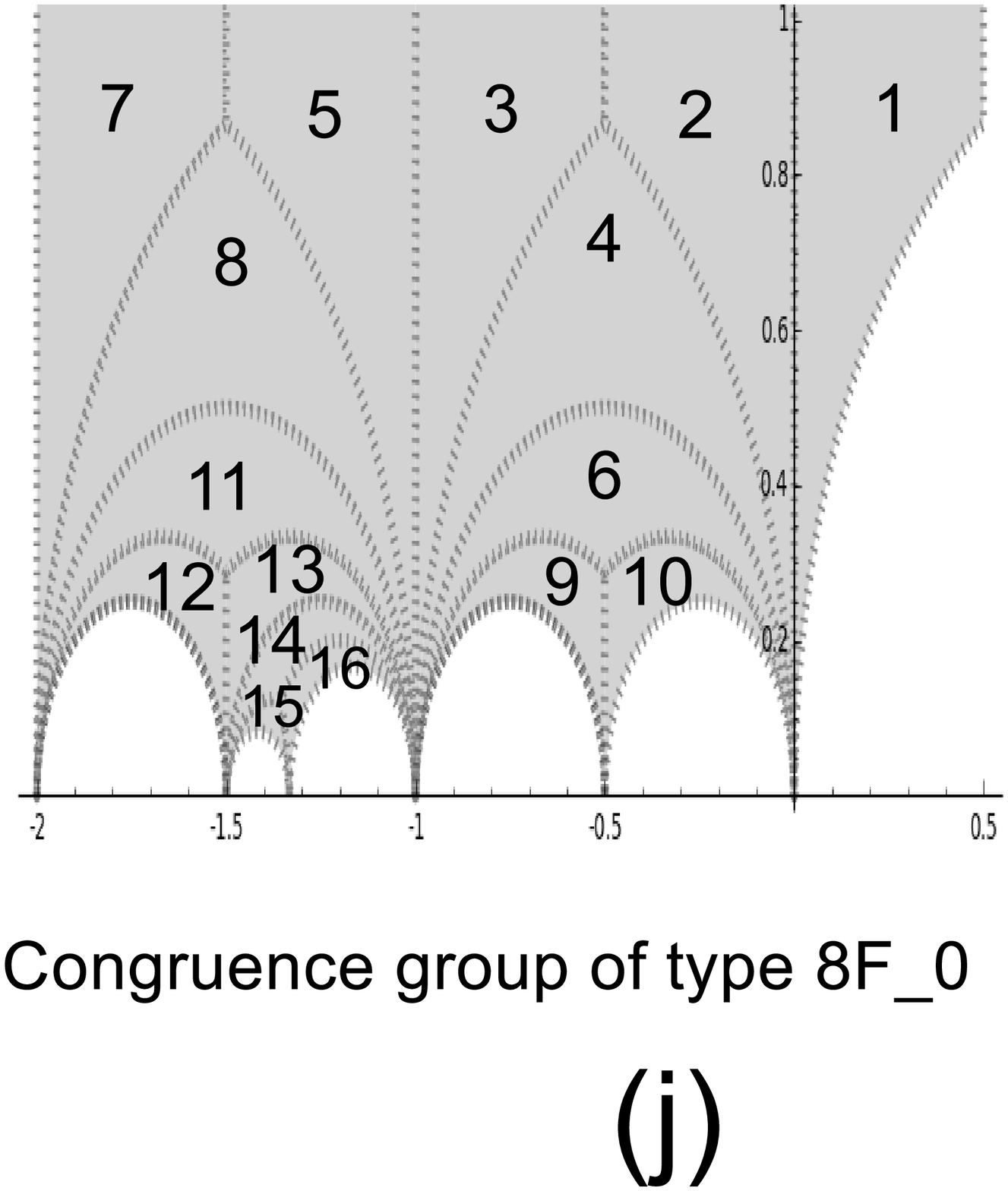}
\includegraphics[width=4cm]{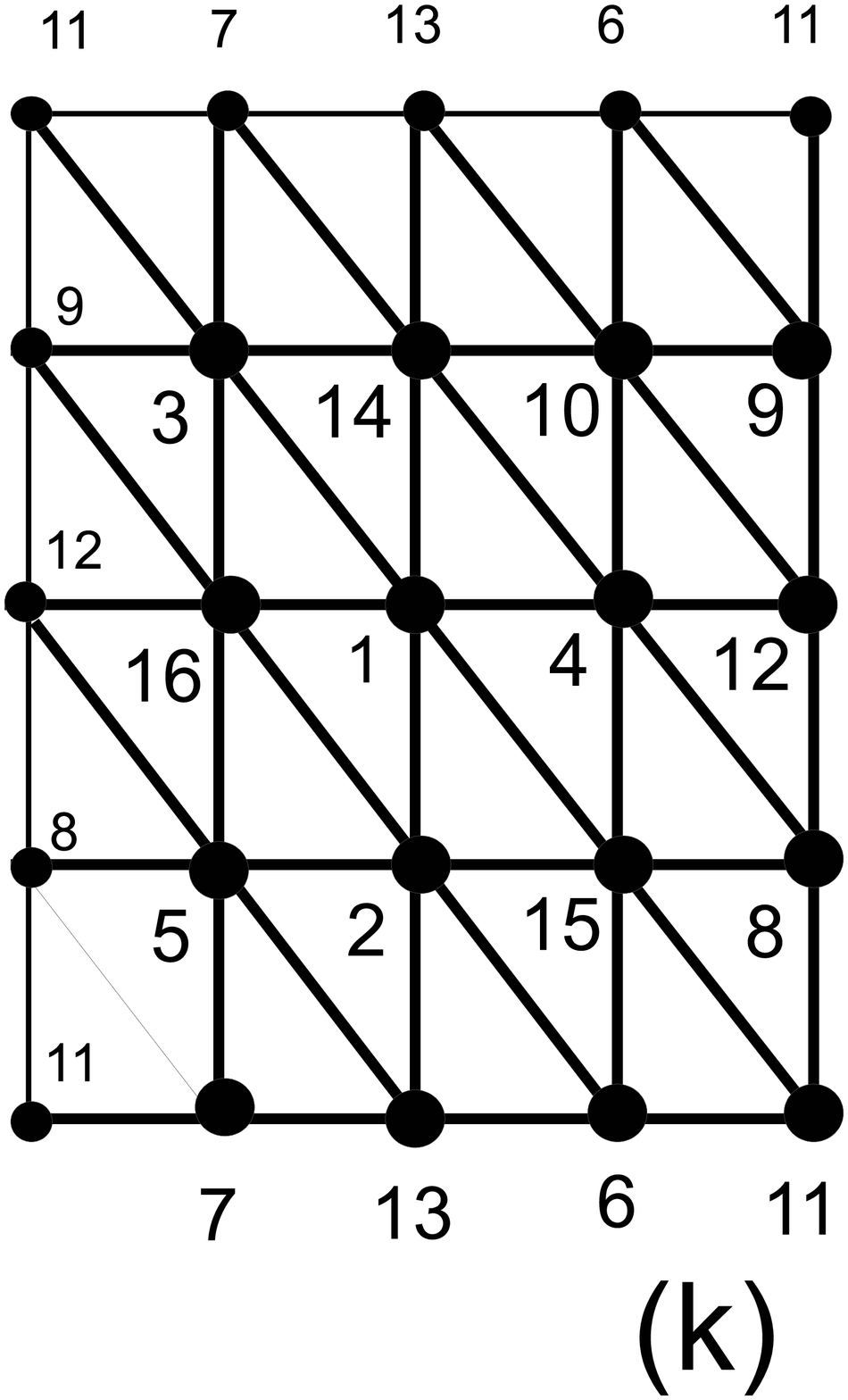}
\caption{The dessin (i) for the Shrikhande graph (k) and the corresponding modular polygon (j). White points are implicit in (i).}
\label{fig6}
\end{figure}

{\bf For index $18$}, the tripartite graph $K(6,6,6)$ may be stabilized (not shown).

\begin{figure}
\centering 
\includegraphics[width=6cm]{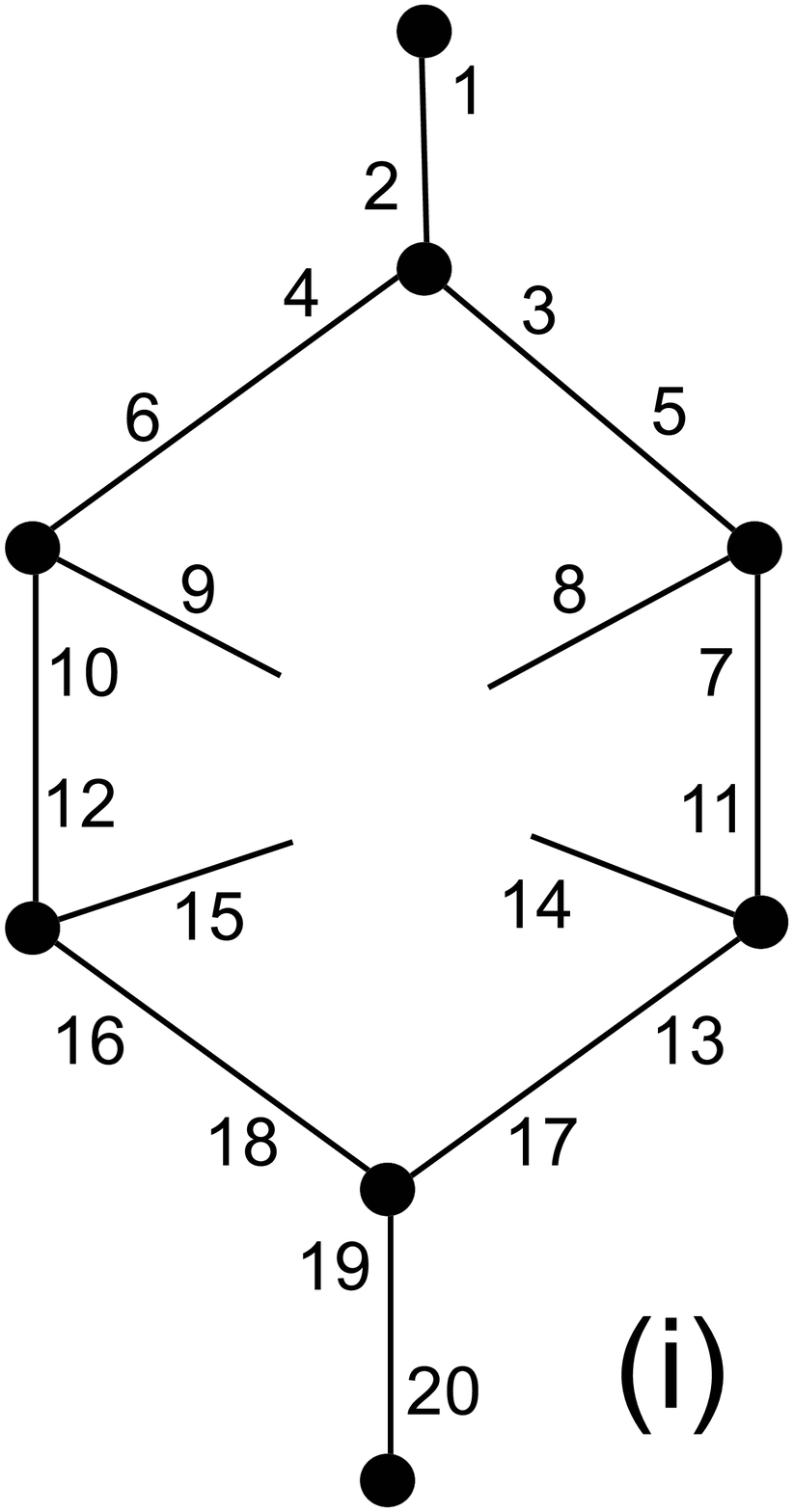}
\includegraphics[width=6cm]{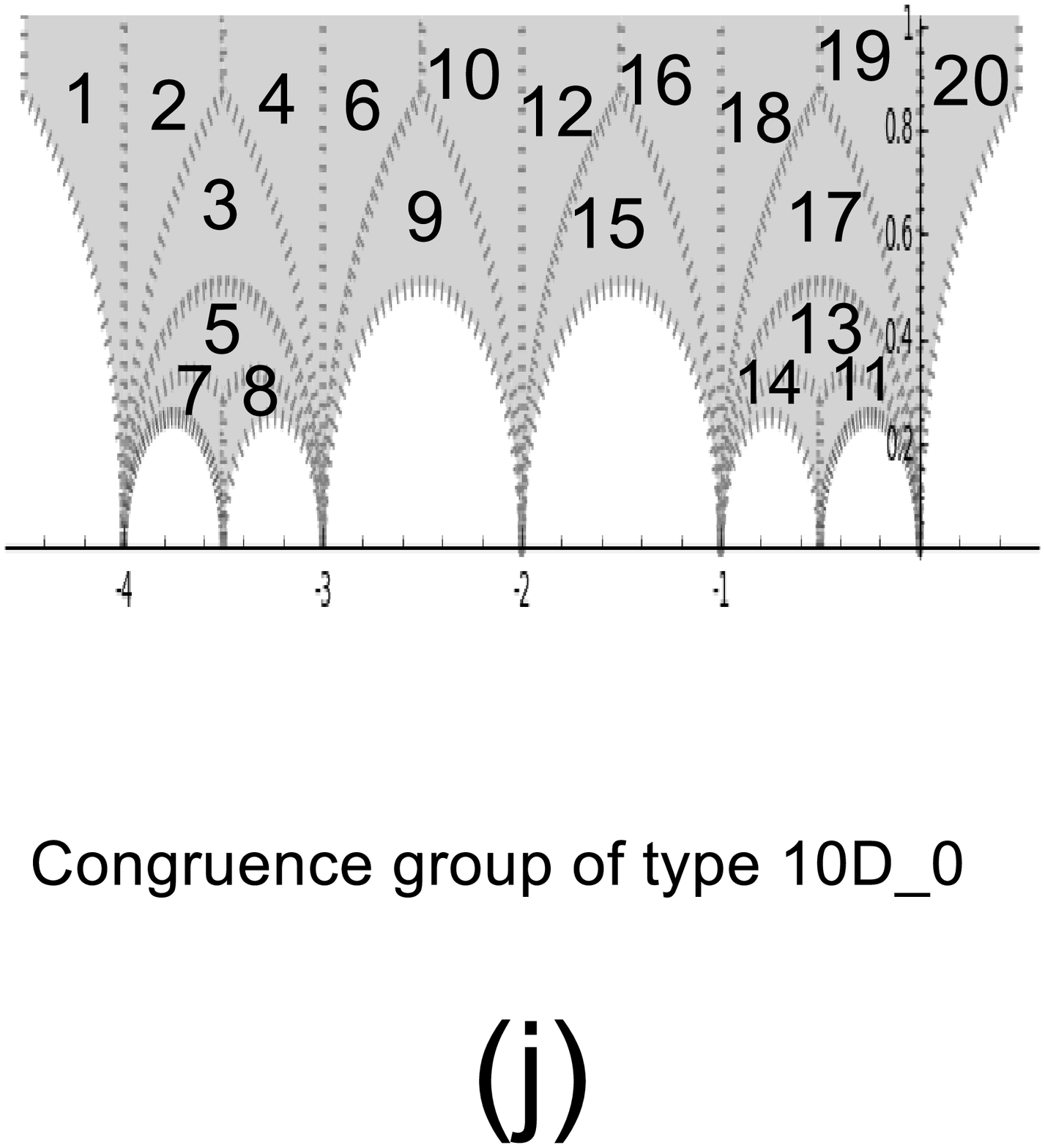}
\includegraphics[width=7cm]{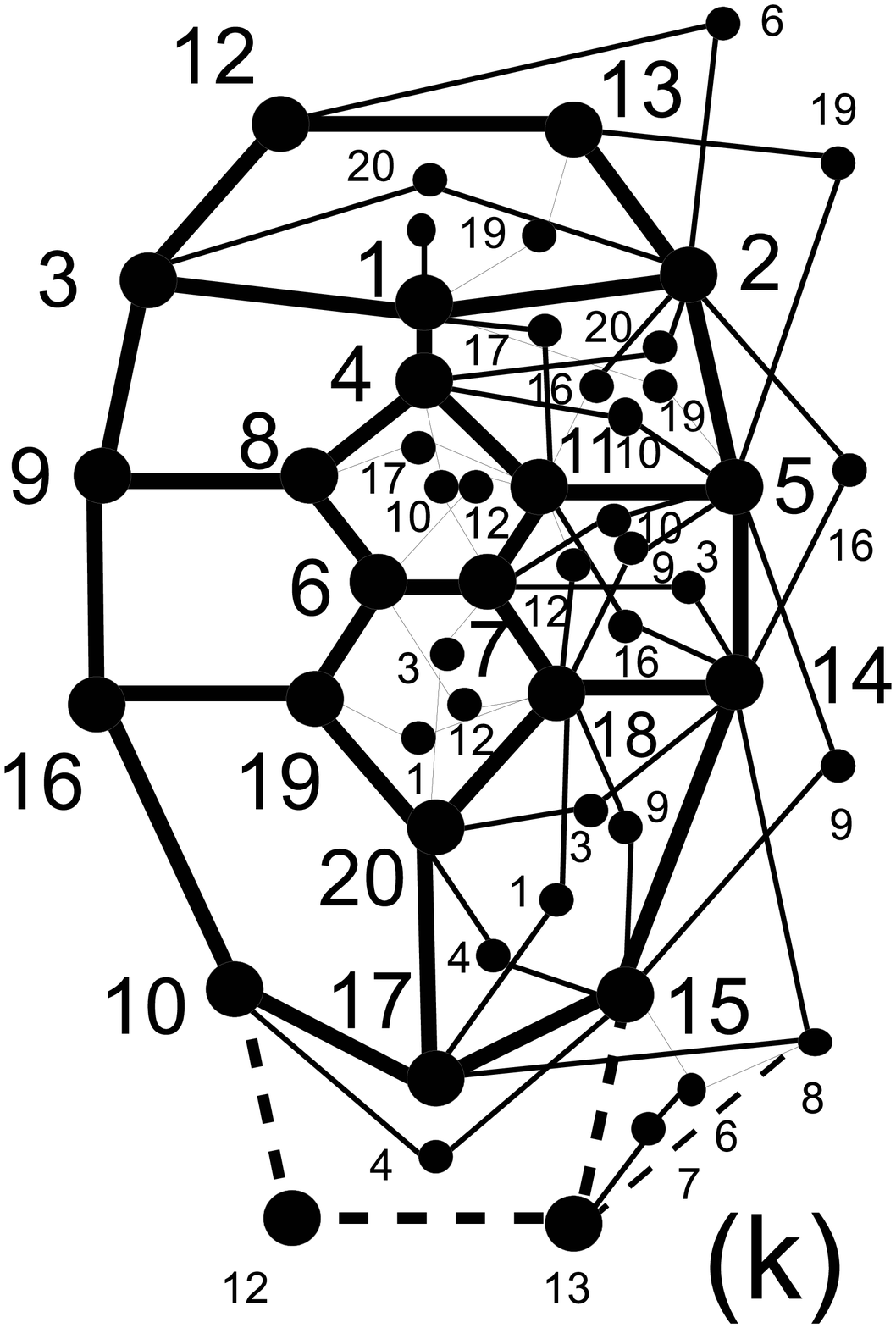}
\caption{The dessin (i) and the corresponding modular polygon (j) stabilizing a $6$-regular graph on $20$ vertices and $60$ edges shown in (k). White points are implicit in (i). Since there is a vertical symmetry axis in (k), only half of the geometry is fully shown.}
\label{fig7}
\end{figure}

\begin{figure}
\centering 
\includegraphics[width=4cm]{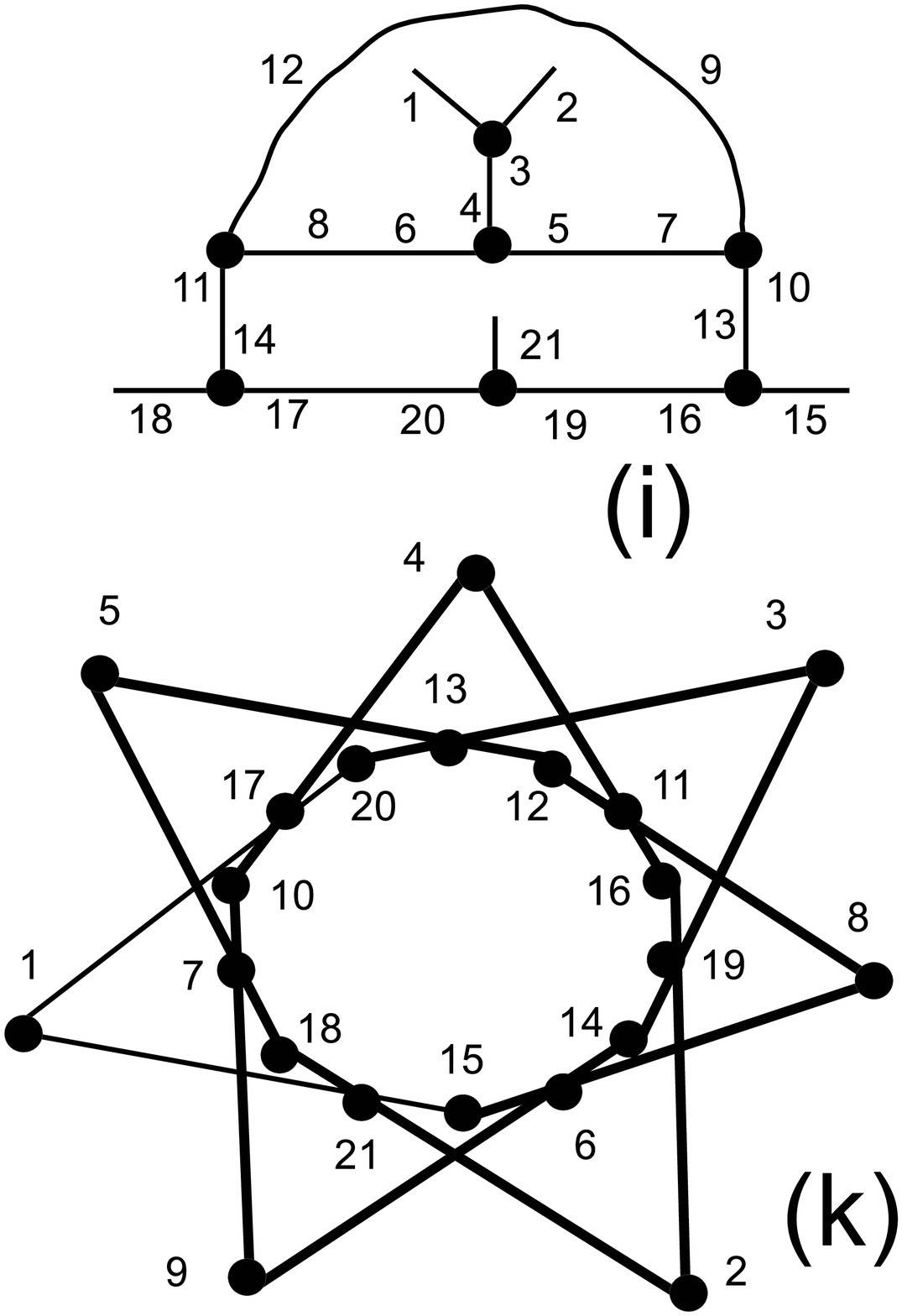}
\includegraphics[width=4cm]{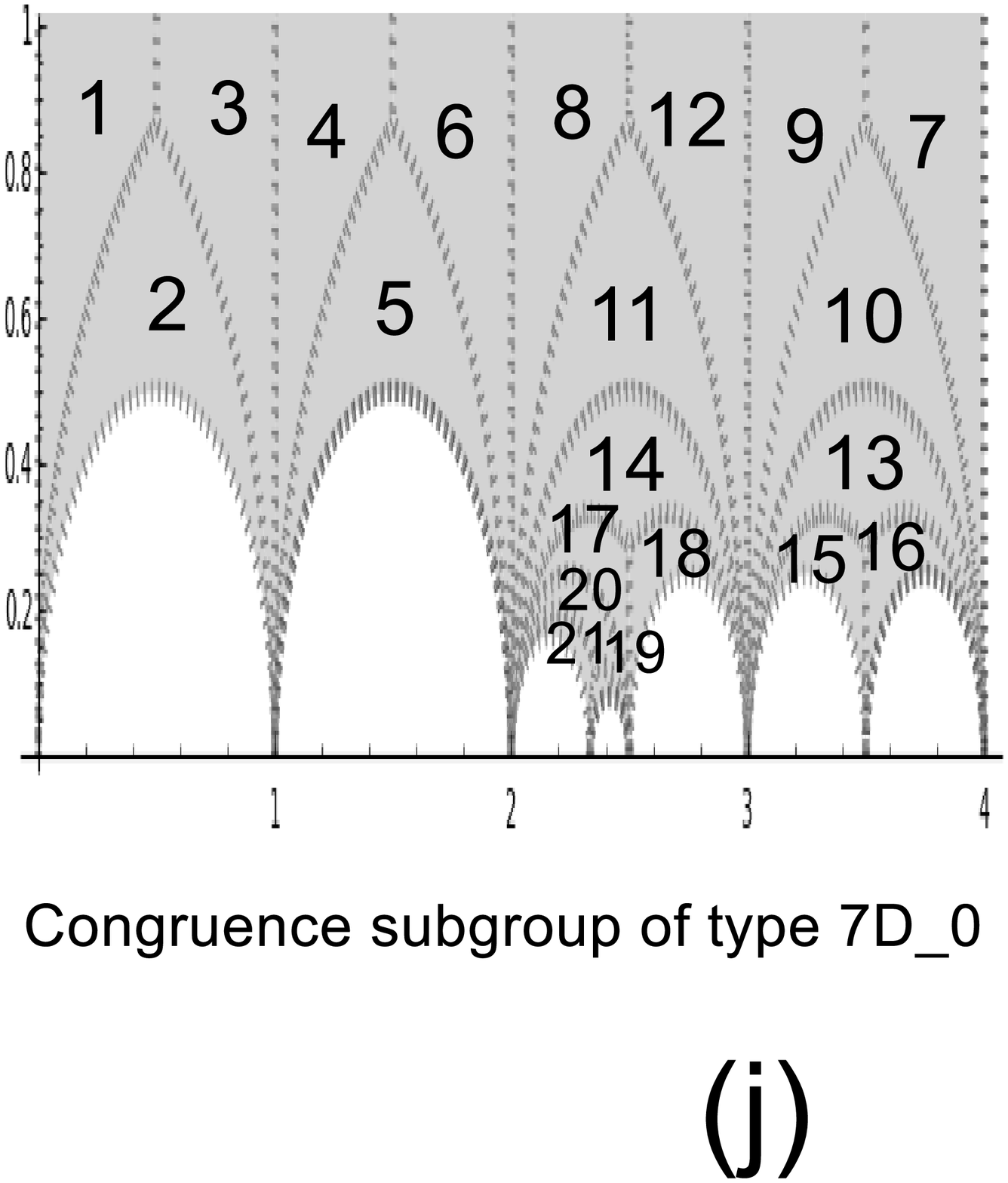}
\caption{The dessin (i) for the thin generalized hexagon $GH(2,1)$ shown in (k), and the corresponding modular polygon (j). White points are implicit in (i).}
\label{fig8}
\end{figure}

{\bf For index $20$}, using the finite representation of the modular group $\Gamma$, the smallest permutation groups is isomorphic to the symmetric group $S_5$, of order $120$. It stabilizes a {\bf highly symmetric $6$-regular graph with $60$ edges}, of spectrum $[6^1,2^5,0^{10},-4^4]$ and of automorphism group isomorphic to $2^{10} \rtimes S_5$. The graph is triangle free, with $135$ squares, $384$ pentagons, $640$ hexagons and no heptagon. In Fig. 7k, the graph is pictured as a set of $10$ pentagons covering the $20$ vertices and a set of (uniquely defined) ordinary quadrangles containing two adjacent edges. The corresponding dessin is shown in Fig. 7i and its corresponding drawing in the upper-half plane is in Fig. 7j. It corresponds to the congruence subgroup of type $10D_0$ in \cite{Cummins} that has level $10$, number of elliptic points $(\nu_2,\nu_3)=(4,2)$ and cusps of structure $10^2$.

{\bf For index $21$}, using the finite representation of the modular group $\Gamma$, the smallest permutation groups are of order $126$ and $168$.
For order $126$ (two cases), the multipartite group $K(3,3,3,3,3,3,3)$ is stabilized (not shown). There is a unique group isomorphic to PSL(2,7), of order 168, that leads to the dessin shown in Fig. 8i and its corresponding drawing in the upper-half plane as in Fig. 8j. It corresponds to the congruence subgroup of type $7D_0$ in \cite{Cummins} that has level $7$, number of elliptic points $(\nu_2,\nu_3)=(5,0)$ and cusps of structure $7^3$. The dessin (i) stabilizes the {\bf thin generalized hexagon $GH(2,1)$ with $14$ lines} shown in Fig. 8k. Observe that the (non modular) dessin shown in Fig. 6a of \cite{PlanatMM2016} is not filtered by our process.

{\bf For index 24}, the finite representation of $\Gamma$ allows to stabilize the multipartite graphs $K(8,8,8)$ and $K(6,6,6,6)$, as well as the rank-$6$ configuration $[24_{24},192_3]_{(6)}$. It arises from a dessin d'enfant of genus $1$ whose type is 12$F_1$
in \cite{Cummins} (level 12, $(\nu_2,\nu_3)=(0,0)$ and cusp structure $[2^1,4^1,6^1,12^1]$).

{\bf For index 25}, the finite representation of $\Gamma$ stabilizes a remarkable configuration $[25_3, 15_5]_{(7)}$ given in Fig. 9k whose collinearity graph is associated to the {\bf orthogonal array} $OA(5,3)$ \cite{Brouwer}. The dessin (i) corresponds to the modular polygon (j). The latter is not a congruence subgroup of $\Gamma$, it has level $10$, $(\nu_2,\nu_3)=(5,1)$ and cusp structure $[10^2,5^1]$.

\begin{figure}
\centering 
\includegraphics[width=4cm]{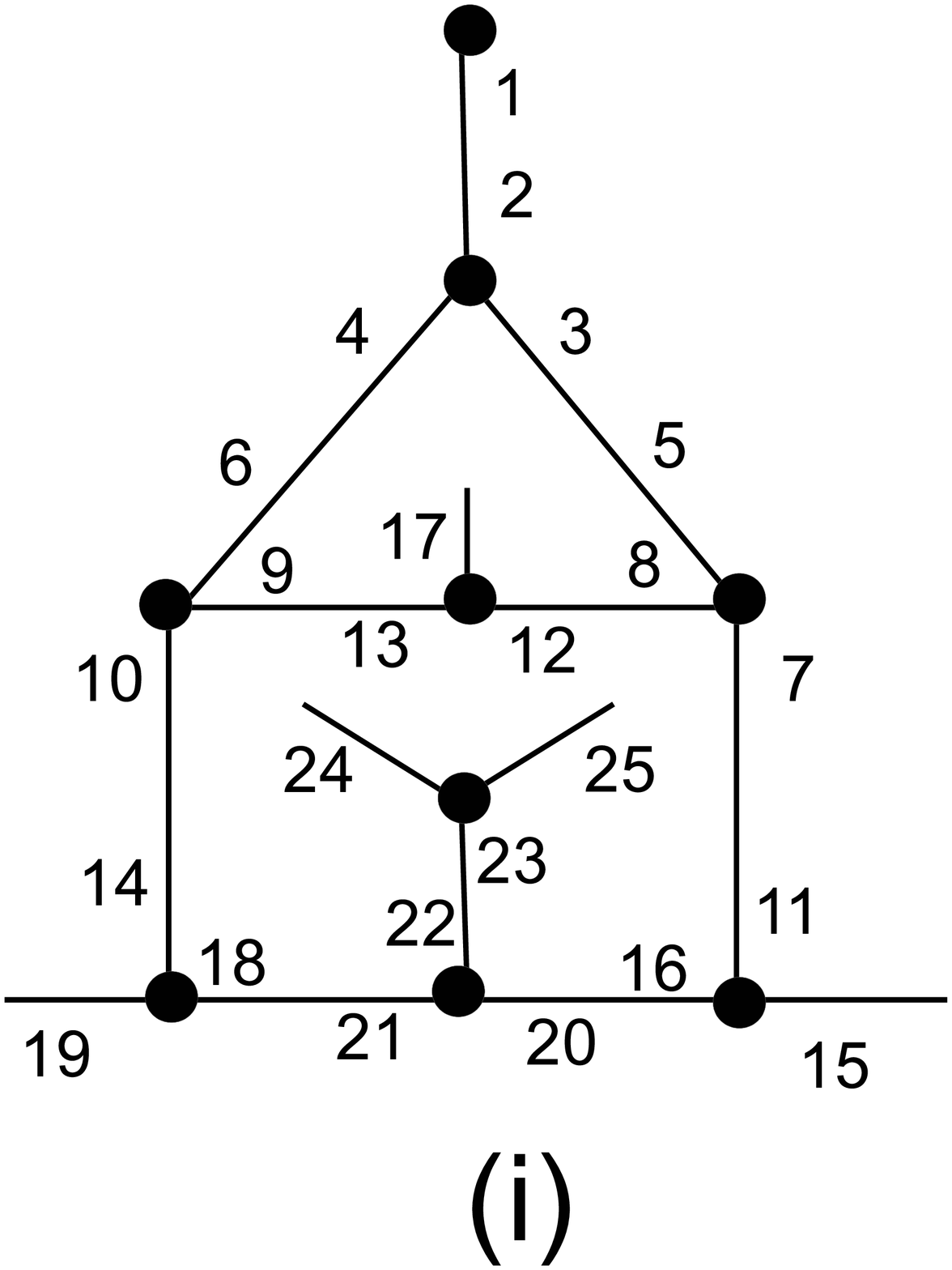}
\includegraphics[width=4cm]{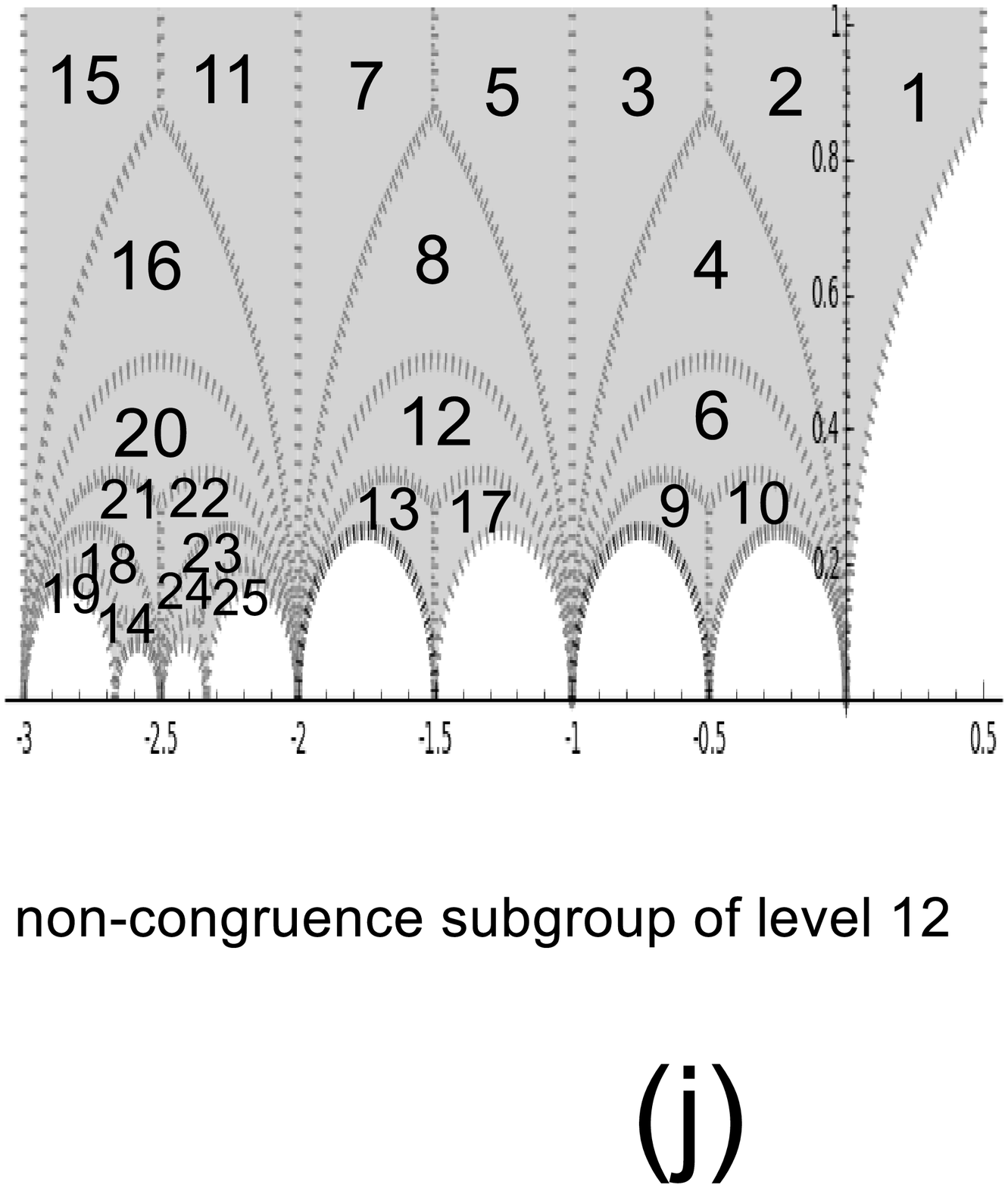}
\includegraphics[width=4cm]{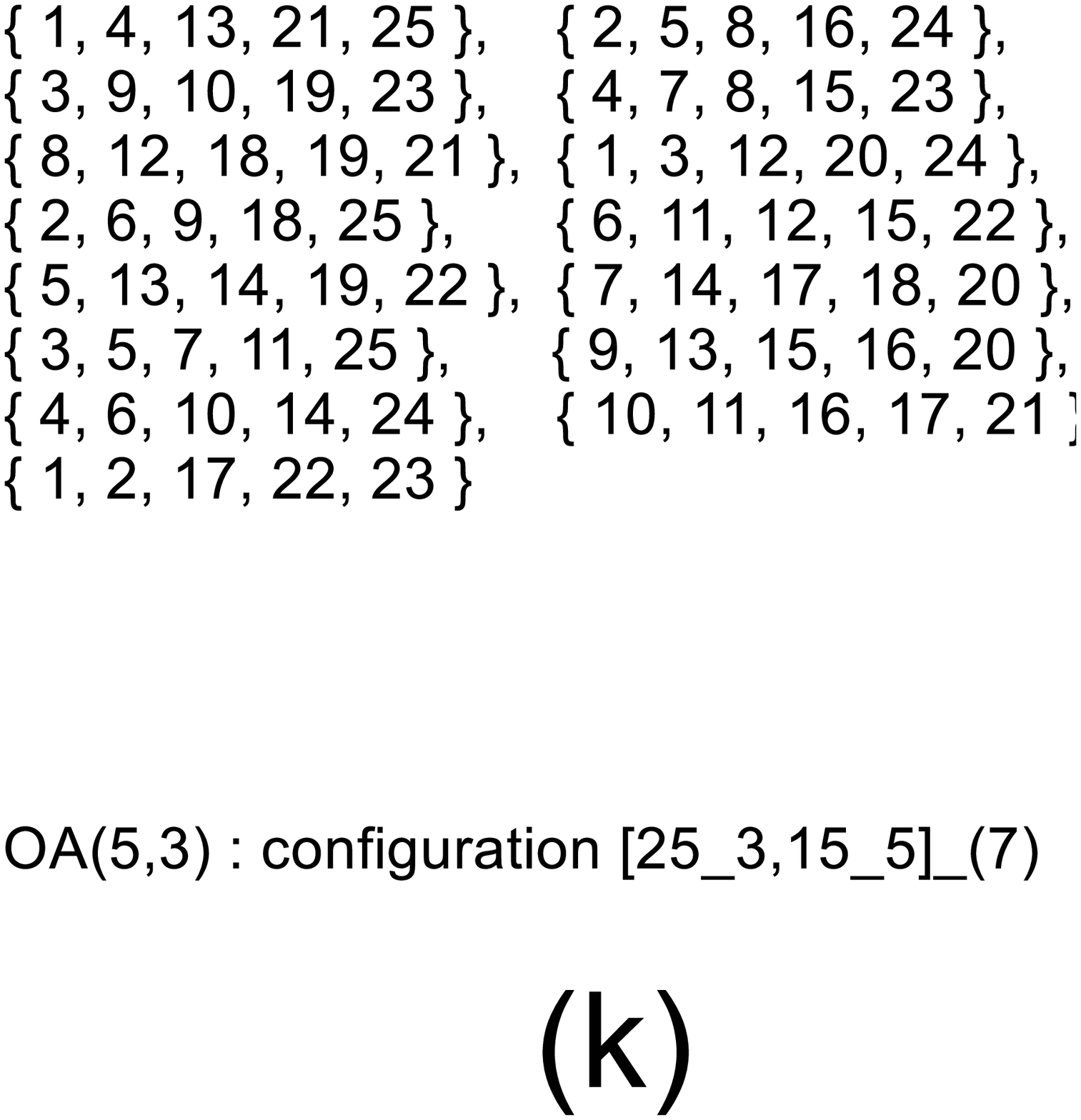}
\caption{The dessin (i) and the corresponding modular polygon (j) corresponding to the configuration of the orthogonal array $OA(4,3)$.}
\label{fig9}
\end{figure}

\begin{figure}
\centering 
\includegraphics[width=4cm]{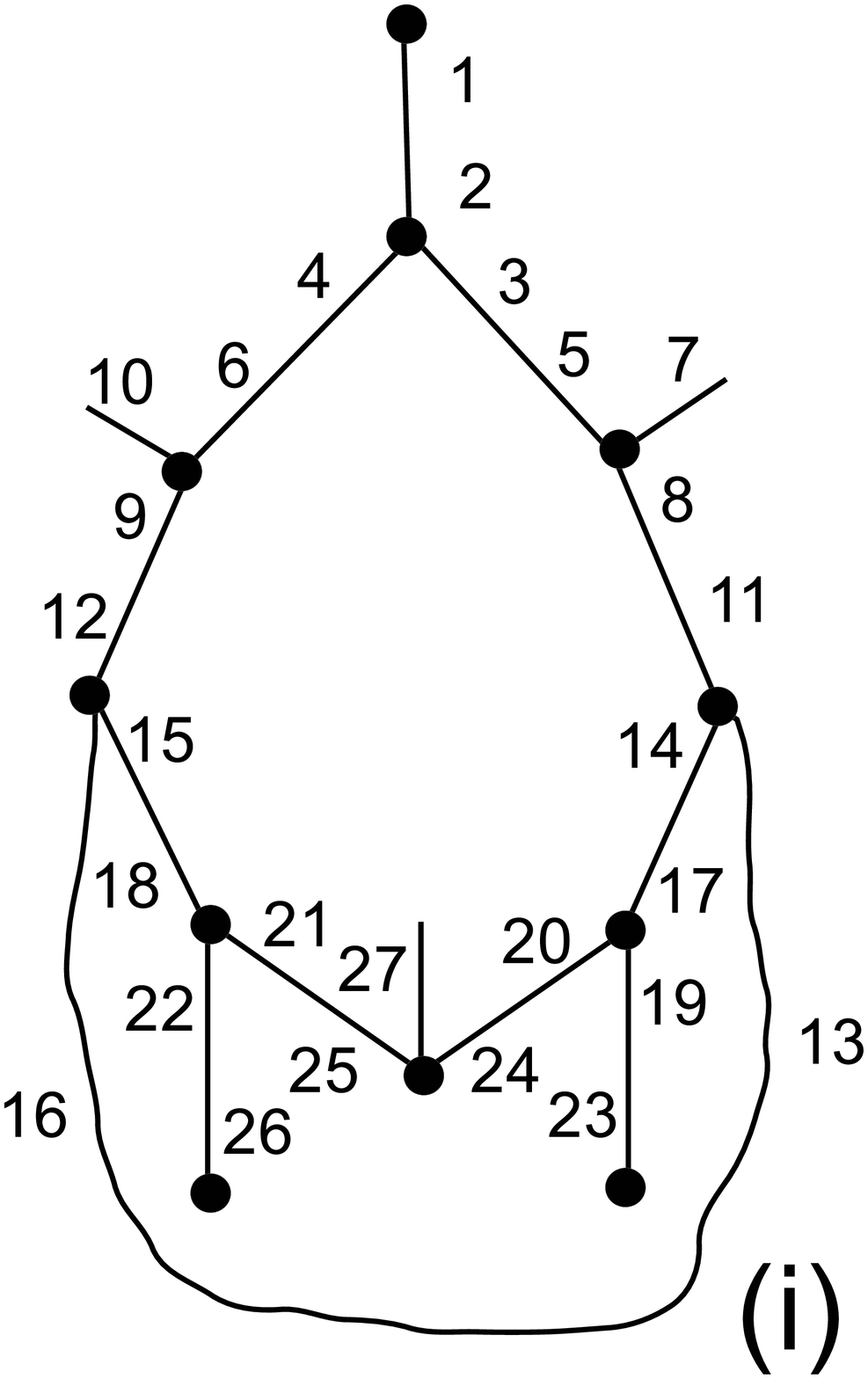}
\includegraphics[width=4cm]{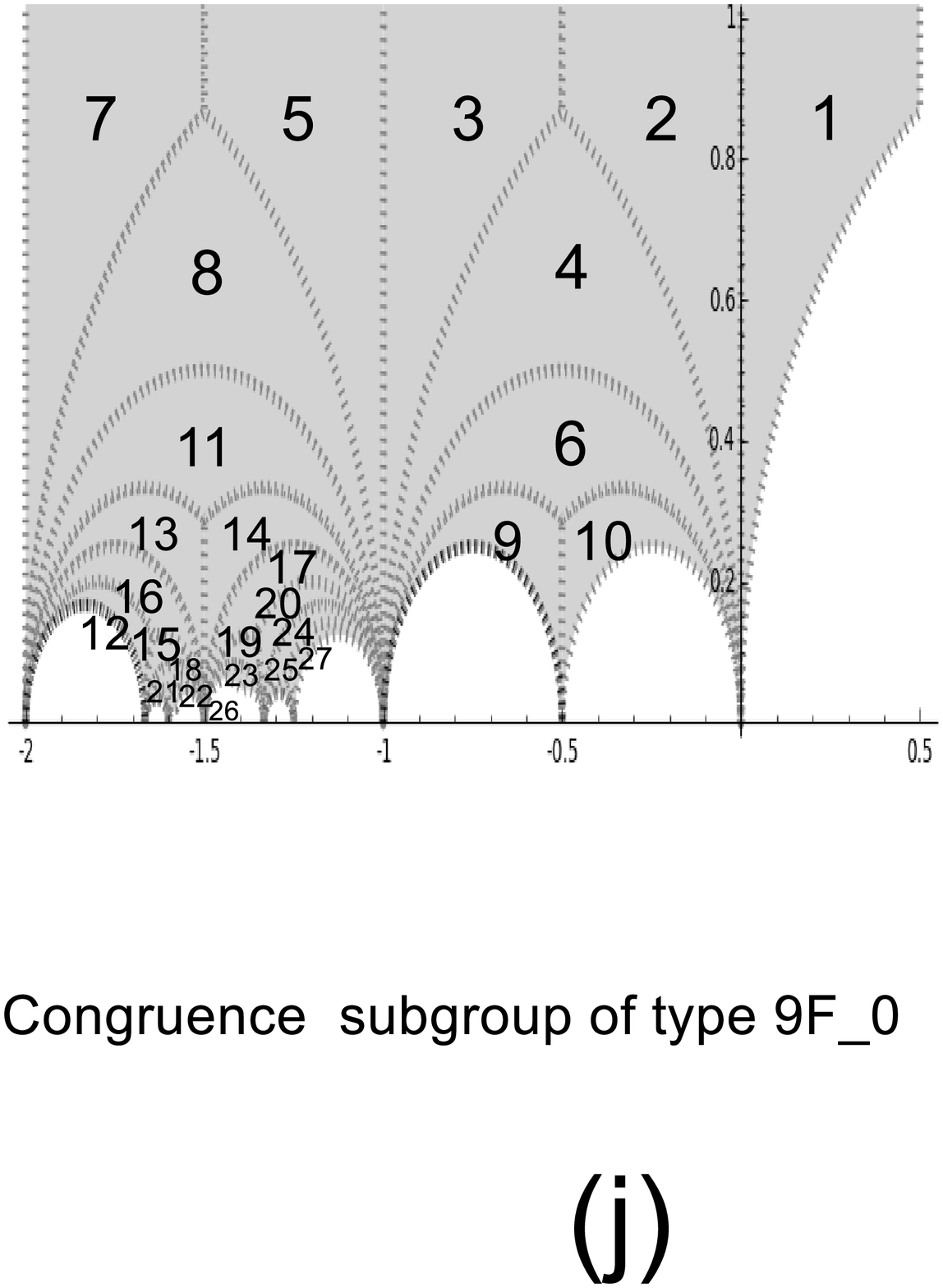}
\includegraphics[width=4cm]{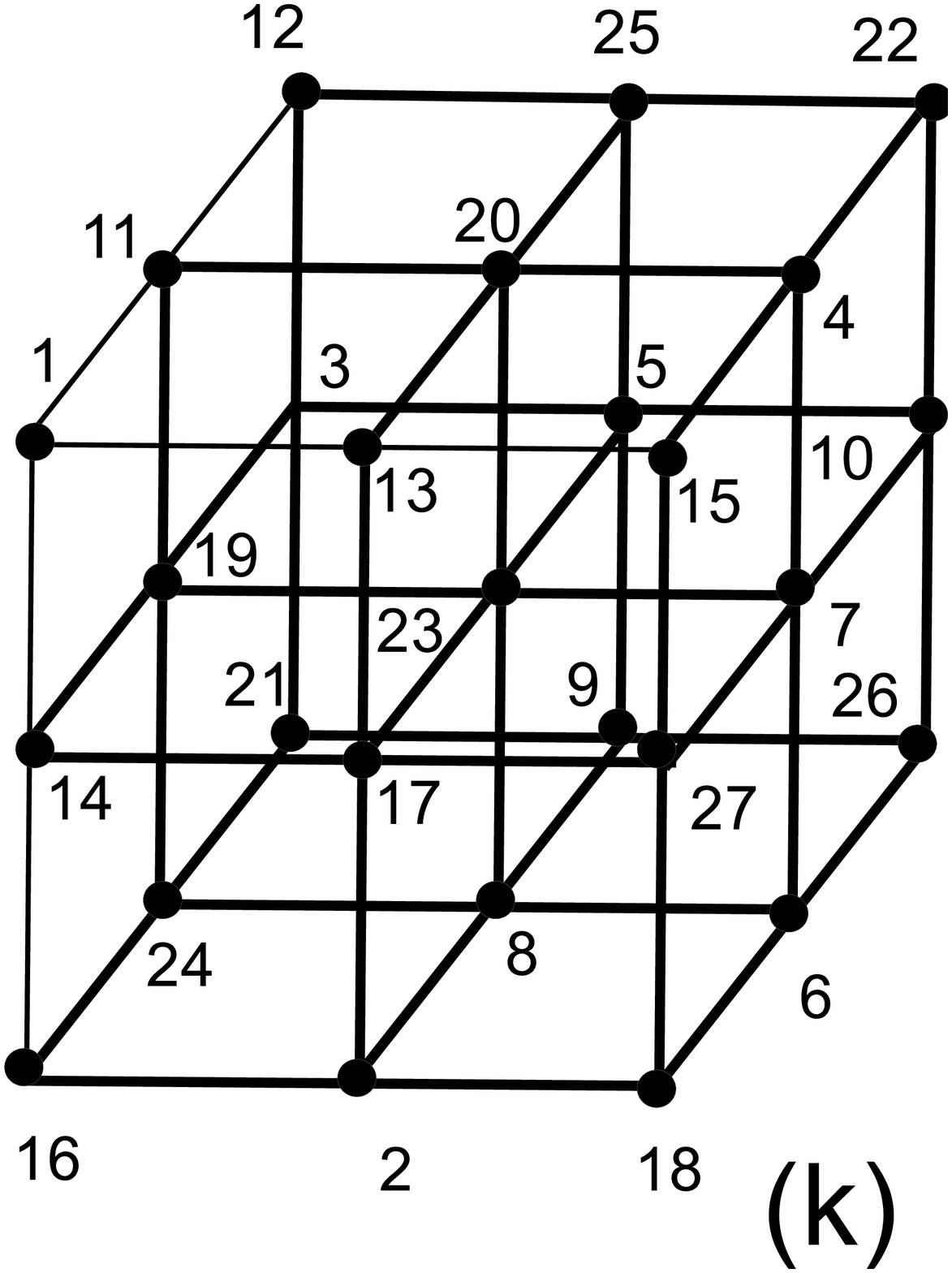}
\caption{The dessin (i) and the corresponding modular polygon (j), of type $9F_0$, stabilizing the smallest slim dense near hexagon (alias the $3\times 3\times 3$ grid) $[27_3]_{(5)}$ shown in (k).}
\label{fig10}
\end{figure}

{\bf For index 27}, the finite representation of $\Gamma$ stabilizes the multipartite graphs $K(3,3,3,3,3,3,3,3,3)$ with $|P|=162$ and $K(9,9,9)$ with $|P|=486$, as well as a configuration $[27_{56},504_3]_{(4)}$ with $|P|=432$. But the most remarkable configuration, obtained with $|P|=324$, is the configuration $[27_3]_{(5)}$ of the {\bf smallest slim dense near hexagon} (alias the $(3\times 3 \times 3)$-grid \cite{Saniga2010}]), as shown in Fig. 10. The corresponding modular polygon in (b) is of type $9F_0$ in \cite{Cummins} (level $9$, $\nu_2=\nu_3=3$ and cusp structure $[9^3]$).

\section{Contextuality of finite simple groups}

In this section, we make use of some finite representations available from the Atlas of finite group representations \cite{Atlas}. Again we restrict to such groups 
satisfying our Definition 1 for geometric contextuality. An expanded list can be found in \cite{PlanatZoology2016} where a less restrictive view of contextuality is taken into account. In \cite{PlanatZoology2016}, the contextuality parameter $\kappa$ is defined as the number of defective edges (edges with non commuting cosets as their coordinates) over the whole number of edges. In the present paper, we restrict to such configurations whose lines are defined by the same two-point stabilizer subgroup and the contextuality parameter $\kappa$ is defined as the number of lines with non commuting cosets over the whole number of lines. Our list is not exhaustive since it happens to be difficult to check the constraint of Definition 1 for configurations with many points on a line and many lines.

\begin{table}[t]
\begin{center}
% \vspace*{-0.4cm}
\begin{tabular}{|l|c|r|c|r|}
\hline 
group & index & configuration & names & $\kappa$ \\
\hline
$S_4'(2)$ & 15 & $[15_3]_{(3)}$ & $GQ(2,2)$: 2QB & 0.800\\
 & 30 & $[30_{16},160_3]_{(7)}$ &  & 0.900\\
\hline
$S_4(3)$ & 27 & $[27_5,45_3]_{(3)}$ & $GQ(2,4)$ & 0.867\\
 & 40 & $[40_4]_{(3)}$ & $GQ(3,3)$: 2QT & 0.900\\
\hline
$S_6(2)$ & 63 & $[63_{15},135_7]_{(3)}$ & $W_5(2)$: 3 QB & *\\
& 120 & $[120_{28},1120_3]_{(3)}$ &  & 0.975\\
& 126 & $[126_{64},2688_3]_{(5)}$ &  & 0.972\\
& 135 & $[135_{7},315_3]_{(4)}$ & DQ(6,2) & 0.978\\
& 315 & $[315_{3},135_7]_{(5)}$ &  & *\\
& 336 & $[336_{10},1120_3]_{(5)}$ &  & 0.991\\
\hline
$U_3(3)$ & 63 & $[63_3]_{(5)}$ & $GH(2,2)$; 3QB & 0.952\\
 & 63 & $[63_3]_{(4)}$ & $GH(2,2)$ dual & 0.936\\
\hline
$U_3(4)$ & 208 & $[208_{6},416_3]_{(5)}$ & config. over $\mathbb{F}_{16}$   & 0.971 \\
\hline
$0_8^+(2):2$ & 120 & $[120_{28},1120_3]_{(3)}$ &$N0^+(8,2)$ & 0.952\\
%\hline
%$U_3(5)$ & 50 & $[50_{7},175_2]_{(3)}$ & Hoffmann-Singleton & \\
\hline
\end{tabular}
\caption{A few contextual configurations arising from the Atlas of finite group representations [11]. The * symbol means that the triangles of the configuration are stabilized (correspond to the same two-point stabilizer subgroup) but not the full lines.}
\end{center}
\end{table}

Our results are in Table 2. It considers such groups of the Atlas where the finite representation is available. The configurations of generalized quadrangles $GQ(2,2)$ and $GQ(3,3)$ correspond the commutation relations of observables in a two-qubit and two-qutrit system (see Fig. 5 for 2QB and Fig. 1 of \cite{PlanatZoology2016} for 2QT), respectively. The sympectic polar space $W_5(2)$ governs the commutation of a three-qubit system. The generalized quadrangle $GQ(2,4)$ is involved in the investigation of black-hole/qubit correspondance \cite{Saniga2010}. The configurations of the generalized hexagon $GH(2,2)$ and its dual also describe the geometry of a three-qubit system, they are pictured in Figs 5-6 of \cite{Planat2015}. For the meaning of the dual quadrangle $DQ(6,2)$ and of $N0^+8,2)$, see \cite{Brouwer}. The other configurations are already discovered in \cite{PlanatZoology2016}.

The configuration $U_3(4)$ over $\mathbb{F}_{16}$ arises from the finite representation of a subgroup of the modular group $\Gamma$. The corresponding modular polygon is a non congruence subgroup of $\Gamma$, of level $13$, genus $6$, with $(\nu_2,\nu_3)=(16,1)$ and cusp structure $[13^{16}]$. It may be related to the construction of the sporadic Suzuki group (see Sec. 3.4 of \cite{PlanatMoonshine2015}).

There is a wealth of symmetries underlying the hyperplane structure of geometric configurations (see \cite{Planat2015} and \cite{Saniga2010}-\cite{Saniga2013} for examples). Let us feature the {\bf hyperplane structure of the $U_3(4)$ configuration}. A basic hyperplane is defined from points of the collinearity graph that are either at minimum or maximal distance from a selected vertex. There are $208$ such hyperplanes. The other hyperplanes may be easily built from Velkamp sums $H \oplus H'$ of the basic hyperplanes $H$ and $H'$, where the set theoretical operation $\oplus$ means the complement of the symmetric difference $(H\cup H')\setminus (H \cap H')$. One finds $10$ distinct classes of hyperplanes totalizing $2^{16}$ hyperplanes, as described in Table 3.

\begin{table}[t]
\begin{center}
\begin{tabular}{||l|r|c|r|c||}
\hline \hline
class & Pts & Lns & Cps  & Type  \\
\hline\hline
%\vspace*{-.30cm}
%&&&&&&&&\\
I   & 80 &  32 &  195    & [ 80; 0, 64, 16, 0, 0, 0, 0]  \\
\hline
{\bf II}  & 88 &  56 &  208    & [ 88; 0, 12, 75, 0, 0, 0, 1]  \\
\hline
III$_1$ & 96 &  80 &  16640  & [ 96; 3, 15, 30, 30, 15, 3, 0] \\
III$_2$    & 96 &  80 &  2496 & [ 96; 10, 5, 25, 50, 0,  1, 5] \\
\hline
IV$_1$  &104 & 104 &  6240     & [104; 5, 4, 20, 40, 31, 0, 4]	\\	
IV$_2$    &104 & 104 &  24000     & [104; 0, 13, 26, 26, 26, 13, 0]	\\
\hline	
V$_1$   &112 & 128 &   8320    & [112; 0, 3, 15, 46, 30, 15, 3]\\
V$_2$    &112 & 128 &   3900    & [112; 0, 0, 28, 32, 32, 16, 4]\\
\hline
VI  &120 & 152 &  3120 &[ 120; 0, 0, 11, 40, 40, 20, 9]\\	
\hline
VII &136 & 200 &   416 &[ 136; 1, 0, 0, 0, 75, 60]\\
\hline
\hline
\end{tabular}
\caption{The $10$ types of hyperplanes of the $U_3(4)$ configuration. Pts is for the number of poins, Lns is for the number of lines and Cps is for the number of copies. The Veldkamp space is the projective space $PG(15,2)$ as for $GH(2,2)$. All hyperplanes follow from Veldkamp sums starting with the type II hyperplane: the latter is defined from points of the graph that are either at minimum or maximal distance. The Type is in the format of \cite{Frohard1994} (where the hyperplanes of $GH(2,2)$ are investigated).}
\end{center}
\end{table}

\section{Conclusion}

Let us quote Niels Bohr {\it What is it that we humans depend on? We depend on our words... Our task is to communicate experience and ideas to others. We must strive continually to extend the scope of our description, but in such a way that our messages do not thereby lose their objective or unambiguous character. We are suspended in language in such a way that we cannot say what is up and what is down. The word "reality" is also a word, a word which we must learn to use correctly.} (Philosophy of Science Vol. 37 (1934), p. 157).

A schematic of our approach based on a Grothendieck's dessin d'enfant (you may have in mind a chinese character for $P_H$) is in Fig. 11.

\begin{figure}[h]
\centering 
\includegraphics[width=4cm]{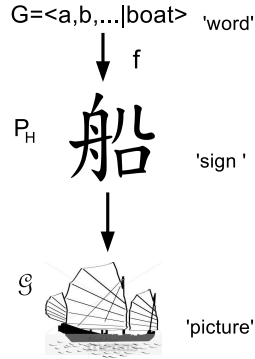}
\caption{A schematic of an object in our theory in terms of two maps, the first one from a word $G$ (through a subgroup $H$ of $G$) to a sign $P_H$ (the dessin d'enfant) and the second one from the sign $P_H$ to the picture $\mathcal{G}$ of the object. }
\label{fig11}
\end{figure}

With two letters from a free group and the right filtering process, we were able to give birth to the geometry of observables governing the quantum experiments with qubits (or qudits) and especially to reveal the basic pieces of the contextuality puzzle, have in mind Mermin's structures with $9$ and $10$ points in Table 1 \cite{Planat2012}-\cite{PlanatSanigaHolweck}. The Definition 1 of contextuality was applied to geometries with a few points (less than 27) mainly those arising from the modular group $\Gamma$ and to a few finite simple groups. In our view, contextuality is highly symmetric and suspended to our language: group theory. We increasing number of points, the geometries we find reveal a contextuality parameter close to its maximal value $1$. Quantum mechanics is contextual as is our language about reality.

%\vspace{12pt}
%\textbf{\large{REFERENCES}}
%\vspace{12pt}

%%%%%%%%%%%%%%%%%%%%%%%%%%%%%%%%%%%%%%%%%%%%%%%%%%%
% TO AUTHORS : PLEASE DO NOT REMOVE THIS LINE BELOW
          %\begin{SCSrefarea} % begins reduced margin area
% REFERENCES AND CITATIONS ARE NOT STANDARD TEX ONES
% SO, PLEASE DO NOT USE THE \begin{thebibliography}
% AND \end{thebibliography} ENVIRONMENT
%%%%%%%%%%%%%%%%%%%%%%%%%%%%%%%%%%%%%%%%%%%%%%%%%%%

%%%%%%%%%%%%%%%%%%%%%%%%%%%%%%%%%%%%%%%%%%%%%%%%%%%
% TO AUTHORS : PLEASE DO NOT REMOVE THIS LINE BELOW
          % \end{SCSrefarea} % end of reduced margin area
%%%%%%%%%%%%%%%%%%%%%%%%%%%%%%%%%%%%%%%%%%%%%%%%%%%

\end{document}